\documentclass[final,authoryear,5p,times,twocolumn]{elsarticle}

\usepackage[latin1]{inputenc}
\usepackage[english]{babel}
\usepackage{amsmath}
\usepackage{amsfonts}
\usepackage{amssymb}
\usepackage{wasysym}
\usepackage{graphicx}
\usepackage{url}
\usepackage{color}

\journal{Icarus}

\begin{document}

\begin{frontmatter}



\author[iaps]{D. Turrini\corref{cor1}}\cortext[cor1]{Corresponding author. \newline \indent{\indent{\emph{Email address}: \url{diego.turrini@iaps.inaf.it}}}}
\author[idg]{V. Svetsov} 
\author[sv]{G. Consolmagno}
\author[nu]{S. Sirono}
\author[bu]{M. Jutzi}

\address[iaps]{Institute of Space Astrophysics and Planetology INAF-IAPS, Via del Fosso del Cavaliere 100, 00133, Rome, Italy.}
\address[idg]{Institute for Dynamics of Geospheres, Russian Academy of Sciences, Leninskiy Prospekt 38-1, Moscow 119334, Russia.}
\address[sv]{Specola Vaticana, V-00120, Vatican City State.}
\address[nu]{Earth and Environmental Sciences, Nagoya University, Tikusa-ku, Furo-cho, Nagoya 464-8601, Japan.}
\address[bu]{Physics Institute, Space Research and Planetary Sciences, Center for Space and Habitability, University of Bern, Sidlerstrasse 5, CH-3012 Bern, Switzerland.}

\title{{\color{black}The late accretion and erosion of Vesta's crust} recorded by eucrites and diogenites as {\color{black}an astrochemical} window into the formation of Jupiter and the early evolution of the Solar System}

\begin{abstract}

The circumsolar disc was the birthplace of both planetesimals and giant planets, yet the details of their formation {\color{black}histories} are as elusive as they are important to understand the origins of the Solar System. For decades the limited thickness of Vesta's basaltic crust, revealed by the link between the asteroid and the howardite-eucrite-diogenite family of meteorites, and its survival to collisional erosion offered an important constraint for the study of these processes. Some results of the Dawn mission, however, {\color{black}cast doubts on our understanding of Vesta's interior composition and of the characteristics of its basaltic crust, weakening this classical constraint}. In this work we investigate the late accretion {\color{black}and erosion} experienced by {\color{black}Vesta's crust} after its differentiation and recorded in the composition of eucrites and diogenites and show that it offers {\color{black}an astrochemical} window into the earliest evolution of the Solar System. {\color{black}In our proof-of-concept} case study {\color{black}focusing on the late accretion and erosion of Vesta's crust} during the growth and migration of Jupiter, the water enrichment of eucrites {\color{black}appears to be} a sensitive function of Jupiter's migration while the enrichment in highly-siderophile elements of diogenites {\color{black}appears to be} particularly sensitive to the size-frequency distribution of the planetesimals. {\color{black}The picture depicted by the enrichments created by {\color{black}late} accretion in eucrites and diogenites is not qualitatively affected by the uncertainty on the primordial mass of Vesta.} {\color{black}Crustal} erosion, instead, is more significantly affected {\color{black}by said uncertainty} and Vesta's crust survival appears to be mainly useful to study violent collisional scenarios {\color{black}where highly energetic impacts can strip significant amounts of vestan material while limitedly contributing to Vesta's late accretion}. {\color{black}While our proof-of-concept case study is based on a simplified physical model and explores only a limited set of scenarios, our results suggest that the astrochemical record of the late accretion and erosion of Vesta's crust provided by eucrites and diogenites can be used as a tool to investigate any process or scenario associated to the evolution of primordial Vesta and of the early Solar System.}

\end{abstract}


\end{frontmatter}


%
\section{Introduction}\label{section-intro}
One of the most challenging tasks in the study of the Solar System is that of disentangling the steps of its formation process that took place during the life of the circumsolar disc, {\color{black}specifically over} the timespan extending from the condensation of the Calcium-Aluminum-rich Inclusions (CAIs) $4568.2^{+0.2}_{-0.4}$ Ma ago, \citep{bouvier2010} to the dissipation of the gas from the disc 4-5 Myr later (\citealt{scott2006,johnson2016,wang2017,kruijer2017}{\color{black}, but values up to 10 Myr are possible based on the comparison with circumstellar discs, see e.g. \citealt{fedele2010}}). Among the most important events that occurred during this timespan are the formation of the planetesimals, the appearance of the giant planets, and their migration due to their interaction with the nebular gas (see  \citealt{morbidelli2016} and references therein).

Our understanding of these three processes, however, has been put under scrutiny by new ideas and scenarios. In particular, various authors have argued that the giant planets formed at locations different from their {\color{black}current} ones and underwent a period of extensive migration during the life of the circumsolar disk (see \citealt{morbidelli2016} and references therein). Such an extensive early migration was shown to be associated with a period of dynamical excitation and orbital remixing of the planetary bodies in the circumsolar disc, with major implications for the evolution of the primordial asteroid belt \citep{walsh2011,obrien2014}. 

\begin{figure*}[t]
\centering
\includegraphics[width=\textwidth]{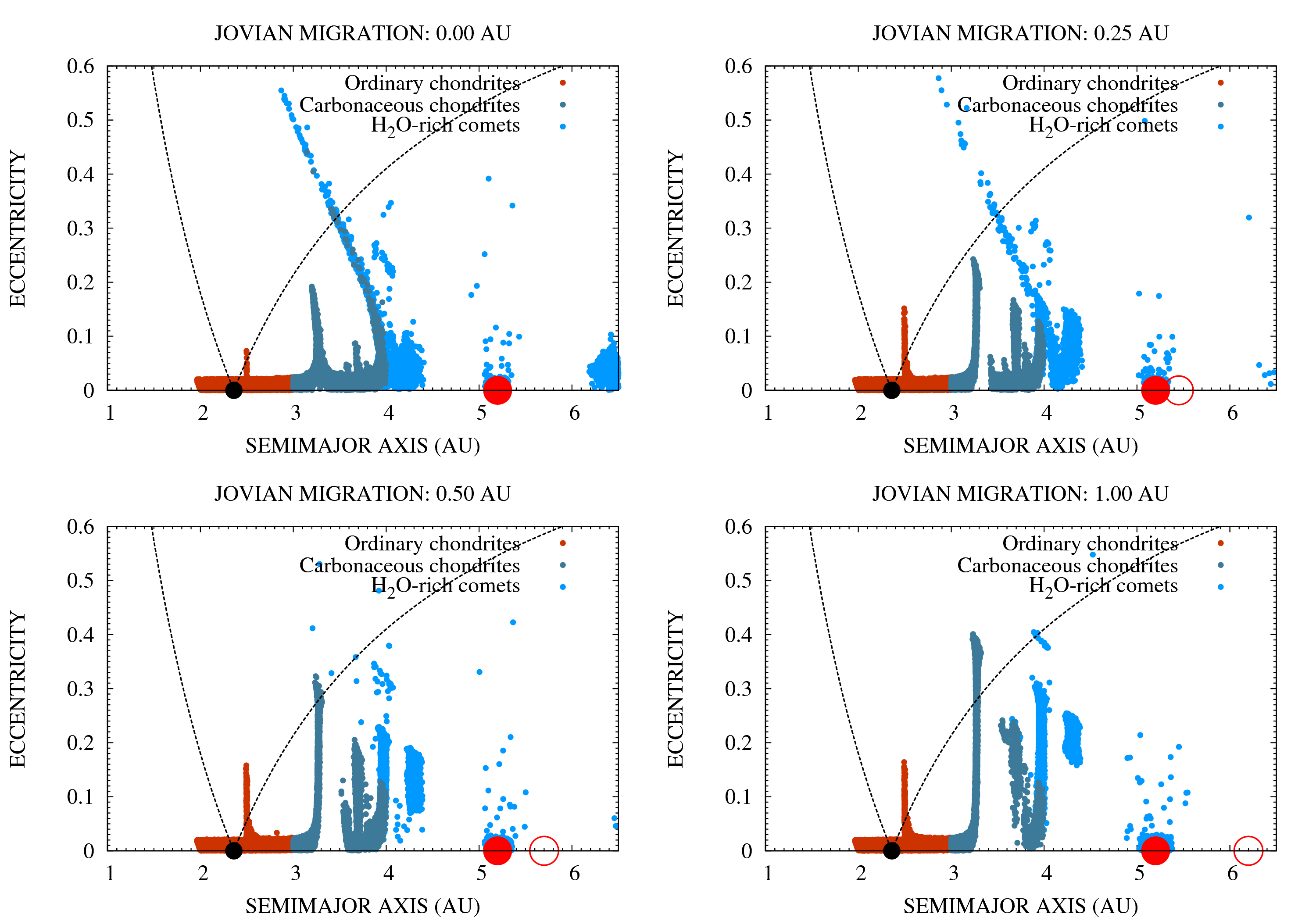}
\caption{Dynamical excitation and radial mixing of the planetesimals in the circumsolar disc in response to Jupiter's mass growth and migration in the simulations by Turrini et al. (2011). The plots show snapshots of the Jovian Early Bombardment 0.2 Myr after the beginning of Jupiter's rapid gas accretion in the four migration scenarios considered by \citet{turrini2011}. The open red circles are the positions of Jupiter at the beginning of the simulations, the bigger red filled ones are the positions of Jupiter once fully formed (see Sect. \ref{section-jupiter}). The smaller black filled circles at 2.36 au mark the orbital position of Vesta. The rocky asteroidal planetesimals analogous to ordinary chondrites that formed between 2 and 3 au are indicated in red (see Sect. \ref{section-planetesimals}). The rocky but water-enriched asteroidal planetesimals analogous to carbonaceous chondrites that formed between 3 and 4 au are indicated in dark cyan (see Sect. \ref{section-planetesimals}). The ice-rich cometary planetesimals that formed beyond 4 au are indicated in blue (see Sect. \ref{section-planetesimals}). Planetesimals inside the region delimited by the two black dotted curves are those that can impact Vesta.}\label{fig-jeb}
\end{figure*}

However, compositional studies of the asteroid belt \citep{demeo2014,	michtchenko2016} disagree on whether an extensive migration of the giant planets is consistent with the current radial distribution of the different kinds of asteroids. On the other hand, the very mass growth of the giant planets was shown to also be capable of {\color{black}triggering phases} of dynamical excitation and radial mixing of the planetesimals even in absence of migration (see Fig. \ref{fig-jeb} and {\color{black}\citealt{turrini2011,turrini2012,turrini2014b,turrini2014c,turrini2015,raymond2017})}. This ambiguity in the early history of the giant planets severely hinders our understanding of the formation of the Solar System.

Most signatures left by these ancient events{\color{black}, like their cratering records,} were removed or altered by the later evolution of the individual planetary bodies or of the Solar System as a whole, making it difficult to  verify conclusively the different models and scenarios (see \citealt{morbidelli2016} and references therein). As our most reliable and temporally resolved source of information on the early life of the Solar System is offered by meteorites, our best chance to solve this conundrum lies in identifying those meteoritic properties that can be linked to the evolution of the nebular environment in which their parent bodies were embedded. 

{\color{black}The aim of this work is to investigate how three specific compositional characteristics of the Howardite-Eucrite-Diogenite (HED) family of basaltic achondritic meteorites and of their parent body asteroid (4) Vesta can be 
jointly used to constrain in a quantitative way the early collisional history of the asteroid and, through that, the dynamical evolution of the circumsolar disc, as first suggested by \citet{turrini2014b} and \citet{turrini2014c}. The three compositional characteristics we will focus on are: the survival of Vesta's basaltic crust, the enrichment in water of eucrites, and the enrichment in highly-siderophile elements of diogenites. 

In exploring the working of the astrochemical constraints provided by these three compositional characteristics, we will consider a proof-of-concept case study focusing on the collisional evolution of primordial Vesta across Jupiter's mass growth in different migration scenarios for the giant planet (the event also labelled as \emph{Jovian Early Bombardment} or JEB, see Fig. \ref{fig-jeb} and \citealt{turrini2011,turrini2012,turrini2014b,turrini2014c,turrini2015}). This case study has been selected as it allows us to reuse previous simulations and results to explore the sensitivity of these astrochemical constraints to a number of physical parameters (namely flux, physical characteristics, size distribution and impact velocity distribution of the impactors and the mass of the primordial Vesta).

The rest of this work is organized as follows. In Sect. \ref{section-context} we will overview the current state of our understanding of asteroid (4) Vesta and of the HEDs. In Sect. \ref{section-constraints} we discuss in more details the compositional characteristics of the HEDs and Vesta we aim to use to constrain the early evolution of the Solar System. In Sect. \ref{section-model} we describe the theoretical tools and the simulations used to in our proof-of-concept case study. Readers interested in the working of the compositional constraints from Vesta and the HEDs can skip this section bearing in mind that, due to the exploratory nature of this work, some of the approximations adopted in the case study will be made for reasons of convenience (e.g. minimizing the need for additional simulations) and will not fit equally well all investigated scenarios. 

The numerical results we will discuss in Sect. \ref{section-results} should therefore be considered only as illustrative of the joint working of the three compositional constraints and the consistency of the investigated scenarios with these compositional constraints will need to be reassessed in more details in future works using more complete physical models. Finally, in Sect. \ref{section-discussion}   we discuss the general application of the compositional constraints from Vesta and the HEDs to other scenarios beyond the simplified ones considered in this work.}

{\color{black}\section{Vesta and the HEDs: witnesses of the beginning}\label{section-context}}

{\color{black}Asteroid (4) Vesta was identified as the possible source of the Howardite-Eucrite-Diogenite (HED) family of basaltic achondritic meteorites more than 40 years ago \citep{mccord1970,consolmagno1977}. The NASA mission Dawn, which explored the asteroid between 2011 and 2012 \citep{russell2012,russell2013}, recently provided a strong confirmation to the proposed Vesta-HED genetic link \citep{desanctis2012,prettyman2012}. Because of this genetic link, the achondritic nature of the HEDs implies that Vesta is a differentiated asteroid that experienced global melting (see e.g. \citealt{greenwood2014,steenstra2016}).

Members of the HEDs family possess some of the oldest formation ages among the meteoritic samples currently available (see e.g. \citealt{scott2007} and \citealt{day2016} and references therein). These ages date the completion of Vesta's differentiation to no later than 3 Myr after the condensation of CAIs \citep{bizzarro2005,schiller2011}. Based on current estimates, this event occurred immediately before the formation of Jupiter and the other giant planets, which is dated between 3 and 5 Myr after CAIs \citep{scott2006,johnson2016,wang2017,kruijer2017}. These data therefore imply that the JEB was most plausibly the first violent collisional event experienced by the partially molten crust of Vesta after the differentiation of the asteroid.

Meteoritic data from the HEDs provide us also indications on the duration of the volcanic resurfacing of Vesta and on the timescale of solidification of its crust after the differentiation process completed (see \citealt{mcsween2011} for a discussion). Specifically, the basaltic eucrites indicate that the outer basaltic crust of Vesta formed over several episodes of magmatism through a solid conductive lid \citep{roszjar2016} that spanned at least 10 Myr \citep{mcsween2011} and possibly up to 35 Myr \citep{roszjar2016}. Thermal and geophysical models suggest that the conductive lid was a few km thick (3-5 km, see e.g. \citealt{formisano2013,tkalcec2013}). 

In parallel, diogenites indicate that the underlying lower crust slowly solidified over tens of Myr (see \citealt{mcsween2011} and references therein). Because of the timing of Jupiter's formation mentioned above (i.e. the first $\sim$2 Myr after Vesta's differentiation) and of the duration of the bombardment it triggered ($\sim$1 Myr, \citealt{turrini2011,turrini2012}), across the JEB both the eucritic and the diogenitic layers were in a partially molten state (see e.g. \citealt{formisano2013,tkalcec2013} for the results of thermal and geophysical models and \citealt{mcsween2011,greenwood2014,steenstra2016, roszjar2016} for the meteoritic evidences).

The most recent compositional models of Vesta combining the information provided by the HEDs {\color{black}(in particular in terms of elemental abundances)} and by the Dawn mission {\color{black}(in particular the survival of Vesta's basaltic crust and the size of Vesta's metallic core, as discussed below)} with astrochemical constraints have eucrites and diogenites as the main components of the upper and lower layers of Vesta's basaltic crust, whose total thickness should range between 20 and 40 km \citep{mandler2013,toplis2013,consolmagno2015}. The astrochemical constraints used in these models implicitly assume a chondritic or solar composition (in terms of relative abundances, not absolute ones) for the major rock-forming elements, in particular the abundant lithophiles Si, Mg, Ca and Al (see \citealt{consolmagno2015} and in particular their Sects. 3.2 {\color{black}, 3.3 and 4.3 for a more detailed discussion of this subject}). 

As all these elements are expected to condense at temperatures greater than 1500 K in the circumsolar disc (see e.g. \citealt{consolmagno2015}), this implicit assumption is expected to hold throughout all but the innermost and hottest region of the circumsolar disc, spanning a fraction of au. According to these compositional models, Vesta's Fe-rich core, which the Dawn mission estimated to possess a radius of 110-140 km \citep{russell2012,ermakov2014}, is overlaid by a mantle composed of harzburgite containing 60-80\% olivine \citep{mandler2013,toplis2013,consolmagno2015}.
 
}
Vesta's differentiated nature and the limited thickness of its crust inferred by the Vesta-HED link made the survival of this crust an important constraint for the study of the evolution of the asteroid belt and the Solar System (see \citealt{davis1985,coradini2011,obrien2011} and references therein, \citealt{turrini2011,broz2013,turrini2014b,turrini2014c,consolmagno2015,pirani2016}). However, some of the very results of the Dawn mission cast doubt on the reliability of the assumption of chondritic bulk composition for the major rock-forming elements {\color{black}of the present-day Vesta} \citep{jutzi2013,clenet2014,consolmagno2015,turrini2016}. 

{\color{black}Specifically, the Dawn mission revealed the existence of two giant, partly overlapping impact basins, named Rheasilvia and Veneneia, in the Southern hemisphere of Vesta \citep{schenk2012} and confirmed the survival of Vesta's crust at all spatial scales, including inside these two giant basins \citep{desanctis2012,ammannito2013,ruesch2014}. Simulations of the formation of both impact basins suggested a total excavation depth of 40-80 km \citep{jutzi2013} and indipendent impact and geologic studies \citep{ivanov2013,ruesch2014} reported an excavation depth of about 30-45 km for the Rheasilvia basin alone, values at odds with the thickness of Vesta's crust reported by the most recent compositional models \citep{mandler2013,toplis2013,consolmagno2015}.} 

{\color{black}More precisely, it has been pointed out} that the lack of olivine signatures inside the two partly overlapping impact basins Rheasilvia and Veneneia {\color{black}and on Rheasilvia's central peak} \citep{jutzi2013,clenet2014,ruesch2014}, Vesta's density profile and the mass balance of its interior structure estimated by Dawn \citep{consolmagno2015}, and the likely exogenous origin of the limited olivine-rich material on Vesta's surface {\color{black}in the Northern hemisphere} \citep{turrini2016} are all inconsistent with the limited thickness of said crust associated with a chondritic bulk composition in terms of the major rock-forming elements \citep{consolmagno2015}. {\color{black}This argues} for a thicker crust of Vesta, which in turns argues for a non-chondritic bulk composition of the present-day asteroid {\color{black}in terms of its major rock-forming elements} \citep{consolmagno2015}.

\citet{consolmagno2015} discussed this apparent mismatch between the information provided by the HEDs and that coming from Dawn and proposed a possible solution, postulating that the asteroid formed from chondritic material and, after differentiating but before solidifying completely, underwent some altering event that changed its bulk composition to its present one. One proposed event that could produce the required alteration would be a grazing collision of a larger primordial Vesta with {\color{black} a body of comparable size} stripping a significant fraction of its mantle while preserving most of its crust \citep{consolmagno2015}. 

Another possibility is that, following the catastrophic disruption of primordial Vesta, the mantle olivine would be more easily fragmented into smaller bits which could be preferentially swept away by {\color{black}gas drag}, leaving larger basaltic fragments to reaccrete onto an intact metallic core \citep{consolmagno2016}.  Other scenarios might be possible, including the existence of many HED parents whose material might have been reaccreted into the asteroid we today call Vesta \citep{consolmagno2015}. {\color{black}Nonetheless, three common traits to all scenarios discussed to date are that primordial Vesta should have been more massive than present-day Vesta, that the altering event is suggested to be linked to impacts, and that the altering event should have occurred  while Vesta was still partially molten or possessed enough radiogenic heat to eliminate any macroporosity created during the alteration in order to fit the constraints posed by Dawn \citep{consolmagno2015}.}

In principle, finding those evolution tracks for the early Solar System that, {\color{black}within this scenario for Vesta's evolution}, can produce the required altering event or collision can offer {\color{black}a substitute} for the classical constraint posed by the survival of Vesta's basaltic crust. However, as the primordial mass of Vesta is currently unconstrained and different evolution tracks can produce the required alteration \citep{consolmagno2015,consolmagno2016}, attempting to study the early evolution of the Solar System using one of these scenarios alone represents an ill-posed problem. {\color{black}What is required, therefore, is a new and general constraint that does not strongly depends on Vesta's primordial mass and that could be applied to all possible scenarios.}

{\color{black}\section{Eucrites and diogenites: astrochemical constraints on the late accretion and erosion of Vesta}\label{section-constraints}}

{\color{black}From the time Vesta differentiated to the moment its crust solidified completely, the eucritic and diogenitic layers were altered by impacts \citep{turrini2011,turrini2012,day2012,turrini2014b,turrini2014c,sarafian2014}. This alteration manifested in two ways. On one hand, impacts removed material from the vestan crust by ejecting part of the mass excavated during the crater formation process at speeds exceeding the ejection velocity of the asteroid. This mass loss process is also known as \emph{cratering erosion} \citep{davis1979}. On the other hand, impacts delivered mass to the vestan crust in the form of the material from the impacting bodies that survives the collision. This mass accretion process is known as \emph{late accretion} or, when specifically referring to the alteration of the crust of planetary bodies by impacts, \emph{late veneer} (see e.g. \citealt{day2016}). From a geologic point of view, in this work we will specifically focus on the late veneer process.

As discussed in Sect. \ref{section-context}, from the meteoritic data supplied by the HEDs we know that Vesta's basaltic crust formed over several magmatic effusive events through a conductive solid lid \citep{roszjar2016} with an estimated thickness of a few km \citep{formisano2013,tkalcec2013}. These effusive events could have been either volcanic (the ``heat-pipe'' mechanism, \citealt{moore2017}) or impact-triggered \citep{turrini2014b,turrini2014c}: the shock wave created by an impact, in fact, damages the surface material at greater depths than those excavated by the crater itself \citep{melosh1989}, therefore creating paths for the magma to reach the surface. During this global effusive resurfacing, the outer layer of Vesta's crust acting as the conductive lid would be in a dynamic equilibrium state, with newer material replacing and pushing downward the older one \citep{moore2017} together with any contaminant delivered by impacts.

As a consequence, the late veneer of the basaltic eucritic layer could span an interval of at least 10 Myr (see \citealt{mcsween2011} and references therein, \citealt{roszjar2016}). During this temporal interval, material delivered to Vesta's surface would contaminate the basaltic eucrites either by direct injection into the melt or by later incorporation into the magma \citep{turrini2014c}. The late veneer of the diogenitic layers should in principle last longer (at least a few tens of Myr, see \citealt{mcsween2011} and references therein), but in order to reach the diogenitic melt the material delivered by later impacts would need to either penetrate thicker layers of solid crust or be pushed at depth by the reprocessing and sinking of the conductive lid.

After the complete solidification of Vesta's crust, impacts would contaminate only the howarditic layer formed by the brecciation of solid eucritic and diogenitic materials (see e.g. \citealt{turrini2014d,turrini2016} for an in-depth discussion of this process on Vesta). Consequently, the composition of eucrites and diogenites records the early collisional evolution of Vesta when the crust of the differentiated asteroid was still partially molten. Since the collisional history of a planetary body is strongly coupled to the evolution of the surrounding environment, the composition of eucrites and diogenites provides constraints on the evolution of the circumsolar disc and the early Solar System.} As we will show in the following, these constraints do not depend on the specific value of the unknown primordial mass of Vesta (see Sect. \ref{section-context} and \citealt{consolmagno2015}) but only on the assumption that the primordial Vesta was characterized by a chondritic bulk composition of the major rock-forming elements.

\subsection{Eucrites, diogenites and mass loss}

For a primordial Vesta with chondritic bulk composition {\color{black}in terms of the major rock-forming elements}, the composition of eucrites and diogenites and, in particular, their abundance in rare earth elements allows one to constraint the fractional thickness of the original vestan crust (see \citealt{consolmagno2015} and references therein). Specifically, {\color{black}based on astrochemical abundances (see e.g. \citealt{lodders2010} and references therein)} \emph{the basaltic crust represented $15-21\%$ of the primordial mass of the asteroid} (see \citealt{consolmagno2015} and references therein). This result is independent on the primordial mass of Vesta and \emph{depends only on the asteroid possessing chondritic bulk composition in terms of its major rock-forming elements at the time of its differentiation} \citep{consolmagno2015}. 

Even if Dawn confirmed the survival of Vesta's crust at all spatial scales \citep{desanctis2012,ammannito2013,ruesch2014}, the historical constraint posed by such survival is weak due to our ignorance of {\color{black}the absolute value of the initial thickness of Vesta's crust (in place of the relative one supplied by astrochemical constraints), of} the original mass of the primordial Vesta and{\color{black}, should it have been larger than that of present Vesta,} of the amount of crustal material that {\color{black}could have been removed by the altering event together with the excess mantle material} \citep{consolmagno2015}.

Until these unknown factors are more precisely quantified, it is difficult to pinpoint the amount of crustal material that can be removed by {\color{black}cratering erosion} without producing an asteroid inconsistent with the present-day Vesta \citep{turrini2014b}. As such, in our {\color{black}proof-of-concept case study} we will limit ourselves to \emph{discuss how the estimated mass losses {\color{black}caused by cratering erosion} compare to this upper bound of $15-21\%$ of the primordial mass of Vesta}.

\subsection{Eucrites and water accretion}

The first piece of the puzzle provided by Vesta's {\color{black}late veneer} is supplied by basaltic eucrites. While Vesta is globally a volatile-depleted body (see \citealt{consolmagno2015} and references therein), the discovery of small apatite crystals in some basaltic eucritic meteorites \citep{sarafian2013} indicates that small quantities of water were present while the eucritic layer was solidifying. {\color{black}While measurements of the D/H ratio in apatites were interpreted as suggestive of a carbonaceous chondritic origin of Vesta's water \citep{sarafian2014,barrett2016}, the results of \citet{hartogh2011} on the D/H ratio of comet 103P/Hartley 2 indicate that comets could also be a compatible source \citep{turrini2014c}. However, an incompatibility with a cometary origin, if confirmed, would allow to reject all scenarios invoking a major role for comets in delivering water to Vesta.}

{\color{black}While the uncertainty associated to such estimates is large, recent work \citep{stephant2016a,stephant2016b,sarafian2017a,sarafian2017b} attempts to constrain quantitatively the amount of water initially present in the eucritic melt. \citet{sarafian2017a,sarafian2017b} report an upper bound to the water content of the eucritic parent melts ranging between 260-1000 $\mu$g/g, i.e. 0.026-0.1 wt\%. Independently,} \citet{stephant2016a,stephant2016b} suggest that water should have represented less than 0.2 wt.\% of the eucritic parent melts. For a primordial Vesta characterized by a chondritic bulk composition, eucrites should represent about 2/3 of the vestan crust and the latter should represent no more than 15-21\% of the vestan mass (see \citealt{consolmagno2015} and references therein). {\color{black}The values estimated by \citet{sarafian2017a,sarafian2017b} and \citet{stephant2016a,stephant2016b} therefore translate in an \emph{upper bound to the water accreted by primordial Vesta of 1-3$\times$10$^{-4}$ the mass of the asteroid}, which we will adopt as our constraint on the maximum amount of water that could be delivered by Vesta's late veneer.}

\subsection{Diogenites and mass accretion}

The second piece of the puzzle provided by Vesta's {\color{black}late veneer} is supplied by diogenites. Specifically, {\color{black}some} diogenites show an over-abundance in highly-siderophile elements (HSEs) with respect to what would be expected following their preferential migration to the vestan core during differentiation  {\color{black}\citep{day2012,dale2012}. While this over-abundance in principle could be explained in different ways (e.g. as the result of variations in the local concentration in the vestan magma, see \citealt{day2016} and references therein), the fact that over-abundances in HSEs are often paired with chondritic elemental ratios of this elements suggests that they result from a late accretion or late veneer of chondritic material (see \citealt{day2016} and references therein). A similar pattern was shown to hold also for the most HSE-enriched eucrites, while eucrites containing low abundances of HSEs presented markedly non-chondritic elemental ratios for these elements (see \citealt{day2016} and references therein, \citealt{dhaliwal2016}).} 

{\color{black}Assuming a chondritic bulk composition for Vesta at the time of this late veneer or accretion, \citet{day2012} associated the measured enrichment to a total accreted chondritic mass of about $1-2\%$ the primordial mass of the asteroid. Because of the uncertainties in this kind of computations and on the amount of chondritic material delivered to the mantle instead of the crust (late accretion vs. late veneer), and because the temporal interval considered in this work (the duration of the bulk of the bombardment triggered by Jupiter's mass growth is $\sim$1 Myr, see \citealt{turrini2011,turrini2012}) is much shorter than the timespan over which diogenites can be altered (at least 10 Myr or more, see above and \citealt{mcsween2011}), we will adopt the range of values estimated by \citet{day2012} as an \emph{upper bound to the total accreted chondritic mass delivered to Vesta by the late veneer, which should therefore not exceed 1-2\% the mass of the asteroid}, keeping in mind that because of said uncertainties the real upper limit could be much lower.}

\section{Modelling Jupiter's formation and Vesta's collisional evolution}\label{section-model}

In this section we provide a synthetic description of the {\color{black} previous results and of the methods and approximations we used in our proof-of-concept case study to model} the collisional evolution of Vesta during the formation and migration of Jupiter, its effects on the eucritic and diogenitic crust {\color{black}and their dependence on different factors}. {\color{black}As mentioned in Sect. \ref{section-intro}, due to the exploratory nature of this work for reasons of convenience we build on the simulations, methods and results of previous studies. As a result, readers should keep in mind that not all the approximations made will adapt equally well to the different cases explored and the numerical results should be considered only as illustrative.

For} more details on the methods and the dynamical simulations used for the computation of the impact probabilities and velocities we refer the readers 
to \citet{turrini2011}, for a more detailed discussion of the collisional model we refer the readers to \citet{turrini2014b} {\color{black}and \citet{turrini2014c},} while for more details on the numerical model used in the impact simulations we refer the readers to \citet{turrini2014c} and \citet{turrini2016}. {\color{black}Readers interested in a more detailed discussion of the dynamical characterization of the asteroidal impactors on Vesta across the formation and migration of Jupiter are referred to \citet{turrini2011} and 	\citet{turrini2014b}, while those interested in the dynamical characterization of the cometary impactors are referred to \citet{turrini2011} and \citet{turrini2014c}.
}

\subsection{Modelling Jupiter's mass growth and migration}\label{section-jupiter}
In this study we used the n-body simulations performed by \citet{turrini2011} and the associated estimates of the impact probabilities on Vesta as the base for our assessment of the erosional and accretional history of primordial Vesta across Jupiter's formation and migration. Those simulations considered a template of the early Solar System composed of the Sun, the forming Jupiter, Vesta and a disk of planetesimals modelled as massless particles, whose dynamical evolution was followed for $2\times10^{6}$ years. {\color{black}From a physical point of view, the starting time of this temporal window should be located between 2 and 4 Myr after the condensation of CAIs to allow for Jupiter to complete its formation between 3 and 5 Myr after CAIs.}

During the first $\tau_{c}=10^{6}$ years of this simulated timespan, Jupiter's core would grow from its initial mass $M_{0}=0.1\,M_{\oplus}$ to the critical mass $M_{c}=15\,M_{\oplus}$ as:
\begin{equation}
 M_{\jupiter}=M_{0}+\left( \frac{e}{e-1}\right)\left(M_{c}-M_{0}\right)\times\left( 1-e^{-t/\tau_{c}} \right)
\end{equation}
where $\tau_{c}$ can be interpreted as the oligarchic growth timescale of Jupiter's core (see e.g. \citealt{dangelo2011} and references therein).

When Jupiter's core reached the critical mass value $M_{c}$, the nebular gas surrounding Jupiter was assumed to rapidly accrete on the planet, whose mass would grow as:
\begin{equation}
 M_{\jupiter}=M_{c}+\left( M_{J} - M_{c}\right)\times\left( 1-e^{-(t-\tau_{c})/\tau_{g}}\right)
\end{equation}
where $M_{J}=317.83\,M_{\oplus}$ is the final and present mass of Jupiter. The e-folding time $\tau_{g}=5\times10^3$ years adopted by \citet{turrini2011} was derived from the hydrodynamical simulations described in \citet{lissauer2009} and \citet{coradini2010}.

In their simulations, \citet{turrini2011} considered four different migration scenarios: $0$ AU (no migration), $0.25$ au, $0.5$ au and $1$ au (see Fig. \ref{fig-jeb}). In their simulations Jupiter always started on circular {\color{black}and planar} orbits and, in those scenarios where migration was included, started migrating inward as soon its core reached the critical mass of $15\,M_\oplus$. {\color{black}This approximation} {\color{black}is equivalent to neglecting the distinction between Type I and Type II migration and starting the migration of the accreting planet as soon the characteristic migration timescale of the forming Jupiter became of the order of $10^{6}$ years (see \citealt{dangelo2011} and references therein). 

Given that the effects on the asteroid belt of the dynamical excitation of the planetesimals triggered by the mass growth of the forming Jupiter are negligible before the gas accretion phase (see \citealt{turrini2011} and \citealt{raymond2017}), from a physical point of view this approximation can be treated as assuming that Jupiter's core started forming farther away and migrated to its initial position due to Type I migration before the beginning of the simulations. Moreover, because of the negligible effects of the forming Jupiter on Vesta before the gas accretion phase, to first order the adopted approximated treatment of Jupiter's mass growth is not in contrast with the shorter timescales and outer formation regions predicted by the so called ``pebble accretion'' scenario \citep{bitsch2015}.} 

{\color{black}After the giant planet begins to migrate, }Jupiter's orbital radius would evolve as:
\begin{equation}
 R_{\jupiter}=R_{0}+\left( R_{J} - R_{0}\right)\times\left( 1-e^{-(t-\tau_{c})/\tau_{r}}\right)
\end{equation}
where $R_{0}$ is Jupiter's orbital radius at the beginning of the simulation, $R_{J}$ is the final orbital radius and $\tau_{r}=5\times10^{3}$ years. {\color{black}The simulations performed by \citet{turrini2011} using a slower migration ($\tau_{r}=2.5\times10^{4}$ years) indicate that the flux of impactors on Vesta is not significantly affected by the migration rate.}

\subsection{Modelling the primordial Vesta}\label{section-primordial-vesta}
In the simulations of \citet{turrini2011}, Vesta was initially placed on a circular, planar orbit with semimajor axis $a_{v}=2.362$ AU. The asteroid was characterized using the best pre-Dawn estimates of its mass ($m_{v}=2.70\times10^{23}$ g, \citealt{michalak2000}) and mean radius ($r_{v}=258$ km, \citealt{thomas1997}), whose values differ by $2-4\%$ from the ones later estimated by the Dawn mission ($2.59\times10^{23}$ g and $262.7$ km respectively, \citealt{russell2012}). 

While these values were reasonable before the arrival of Dawn, the results of \citet{consolmagno2015} suggest that primordial Vesta {\color{black}could have been} more massive (see Sect. \ref{section-intro}). Because of this uncertainty on primordial Vesta's mass {\color{black}and because a precise assessment of the latter is beyond the scope of this work}, we maintained the template of primordial Vesta used by \citet{turrini2011} and took advantage of the link between impact probabilities and diameter of the asteroid to rescale the impact fluxes to a more massive primordial Vesta's {\color{black}and explore how the three compositional constraints offered by Vesta and the HEDs responded to this change}. 

{\color{black}We therefore initially considered a primordial Vesta  characterized by a diameter similar to its current mean one. This allows us to take advantage of the fluxes of impactors on the asteroid estimated by \citet{turrini2011} (see Sect. \ref{section-collisions}). Similarly, in simulating the outcomes of impacts at different impact velocities on Vesta, we characterized the target body with the current diameter and surface gravity of Vesta (see Sect. \ref{section-collisions}). This choice allows us to take advantage of the simulations of rocky impactors on Vesta performed by \citet{turrini2016} and to simulate only the effects of more realistic cometary impactors than those originally considered by \citet{turrini2014c} (see Sect. \ref{section-collisions}).

The probabilistic method used by \citet{turrini2011} to estimate impact fluxes on Vesta links impact probabilities to Vesta's diameter. As long as Vesta's mass is not so large that the gravity of the asteroid significantly enhances its effective cross-section (see \citealt{turrini2011} and references therein), impact fluxes will scale with the diameter of the asteroid. For the impact velocities estimated by \citet{turrini2011}, this condition is satisfied for a primordial Vesta no more massive than a few times the present asteroid. Similarly, both the mass erosion \citep{holsapple2007} and the mass accretion \citep{svetsov2011} efficiencies scale with the surface gravity of the target asteroid, which for a given average density will scale with its diameter. 

This approach allowed us to estimate, to first order, the mass loss and mass accretion experienced by primordial Vesta for different values of its original mass without the need of performing a large number of additional simulations. More details on the parameters describing Vesta in our collisional simulations are provided in Sect. \ref{section-collisions}, while a discussion of the effects of a larger mass of the primordial Vesta on our results is presented in Sect. \ref{section-results} and \ref{section-discussion}.}

\subsection{Modelling the planetesimal disk}\label{section-planetesimals}
The planetesimal disk was modelled by \citet{turrini2011} as a disk of massless particles evolving under the gravitational influence of the Sun, Jupiter and Vesta. The disk of massless particles was composed by $8\times10^{4}$ particles and extended from 2 au to 10 au.  {\color{black}The massless particles initially possessed eccentricity and inclination (in radians) values comprised between 0 and $3\times10^{-2}$ \citep{turrini2011} and were used as dynamical tracers of the evolution of the planetesimal disk, each particle representing a swarm of real planetesimals.}

The number of real planetesimals populating each swarm and their characteristic diameter depend on the adopted size-frequency distribution (SFD) for the planetesimal disk. In this work we considered a total of four SFDs: two for primordial planetesimals and two for collisionally evolved planetesimals. Each pair of SFDs (primordial and collisionally evolved) refers to a specific nebular environment, namely quiescent or turbulent circumsolar disc.

The massless particles where associated to their diameters by means of Monte Carlo methods. Since this procedure was performed while processing the output of the simulations, the latter did not include the effects of gas drag {\color{black}as they are size-dependent. The choice of neglecting the effects of gas drag allowed us to explore the effects of different SFDs on Vesta's crustal late accretion and erosion without the need to perform a large number of computationally expensive n-body simulations.

While computationally convenient, however, this choice is not dynamically accurate, particularly for km-sized planetesimals, as gas drag acts to damp orbital eccentricities and inclinations, diminishing the population of dynamically excited planetesimals. At the same time, the radial drift caused by gas drag brings more planetesimals into the orbital resonances with Jupiter, which appear to play the leading role in producing the population of impactors on Vesta (see \citealt{turrini2011} and Sect. \ref{section-results}). The results of analogous simulations performed by \citet{weidenschilling2001}, \citet{grazier2014} and \citet{raymond2017} indicate that neglecting the effects of gas drag should not alter the results of this study in a qualitative way by cancelling the JEB.}

Differently from the previous studies of \citet{turrini2014b} and \citet{turrini2014c}, all four considered SFDs where associated to a circumsolar disc possessing a dust-to-gas ratio $\xi_{i}=0.005$ inside the water ice condensation line and $\xi_{i}=0.01$ outside{\color{black}(see below for details on the density profiles of the individual discs)}. The water ice condensation line was assumed at 4 au. The mass of solids comprised between 2 and 3 au amounted to about $2\,M_{\oplus}$ {\color{black}for all four SFDs, consistent with the planetesimals having formed within a Minimum Mass Solar Nebula (see also \citealt{morbidelli2009} and \citealt{weidenschilling2011})}.

All planetesimals inside 4 au were assumed to be rocky asteroids with an average density of 2.4 g/cm$^{3}$ (chosen as a compromise between the densities of volatile-poor and volatile-rich asteroids, see \citealt{britt2002,carry2012,turrini2014d} and references therein) while those beyond were assumed to be ice-rich cometary bodies, constituted at $50\%$ of their mass by water ice and at $50\%$ by rock, with an average density of 1  g/cm$^{3}$. Planetesimals formed between 3 and 4 au were assumed to possess $10\%$ of their mass as water in the form of hydrated minerals, similarly to carbonaceous chondrites \citep{jarosewich1990,robert2003}. 

The transition at 3 au, while somewhat arbitrary, is consistent with the current distribution of low albedo volatile-rich asteroids being the result of their inward radial diffusion over the life of the Solar System \citep{michtchenko2016}. Moreover, the flux of impactors on Vesta originating from beyond 3 au is due to the 2:1 resonance with Jupiter (located at 3.3 au or outward depending on the Jovian migration, see Fig. \ref{fig-jeb} and \citealt{turrini2011}), so our analysis is not particularly sensitive to the actual heliocentric distance of this transition.

{\color{black}The four SFDs we considered in our case study are described in more detail in the following. A comparison of the average diameters of the planetesimals as a function of their orbital distance from the Sun for the two primordial SFDs is shown in Fig. \ref{fig-primordial_sfds}, while in Fig. \ref{fig-evolved_sfds} we show the comparison between the two collisionally evolved SFDs in the reference orbital region comprised between 1 and 4 au considered by \citet{weidenschilling2011} and \citet{morbidelli2009} (see Sects. \ref{section-pqce} and \ref{section-ptce} for the discussion of their extension to the orbital region between 4 and 10 au).

\begin{figure}[t]
\centering
\includegraphics[width=\columnwidth]{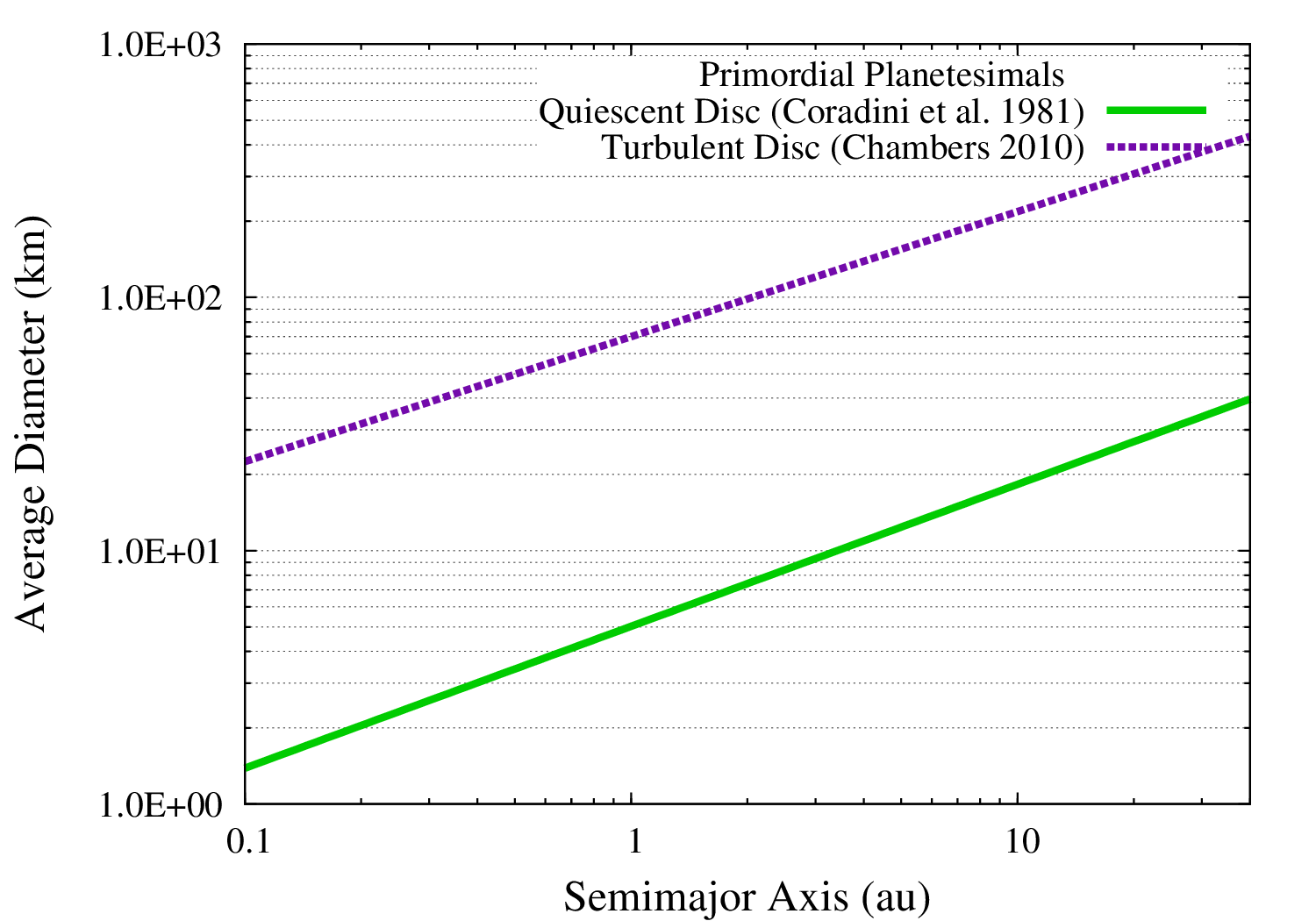}
\caption{Comparison between the average diameters of the planetesimals as a function of their orbital distance from the Sun for the two primordial SFDs considered in our case study (see Sects. \ref{section-pqp} and \ref{section-ptp} for details).}\label{fig-primordial_sfds}
\end{figure}

\begin{figure}[t]
\centering
\includegraphics[width=\columnwidth]{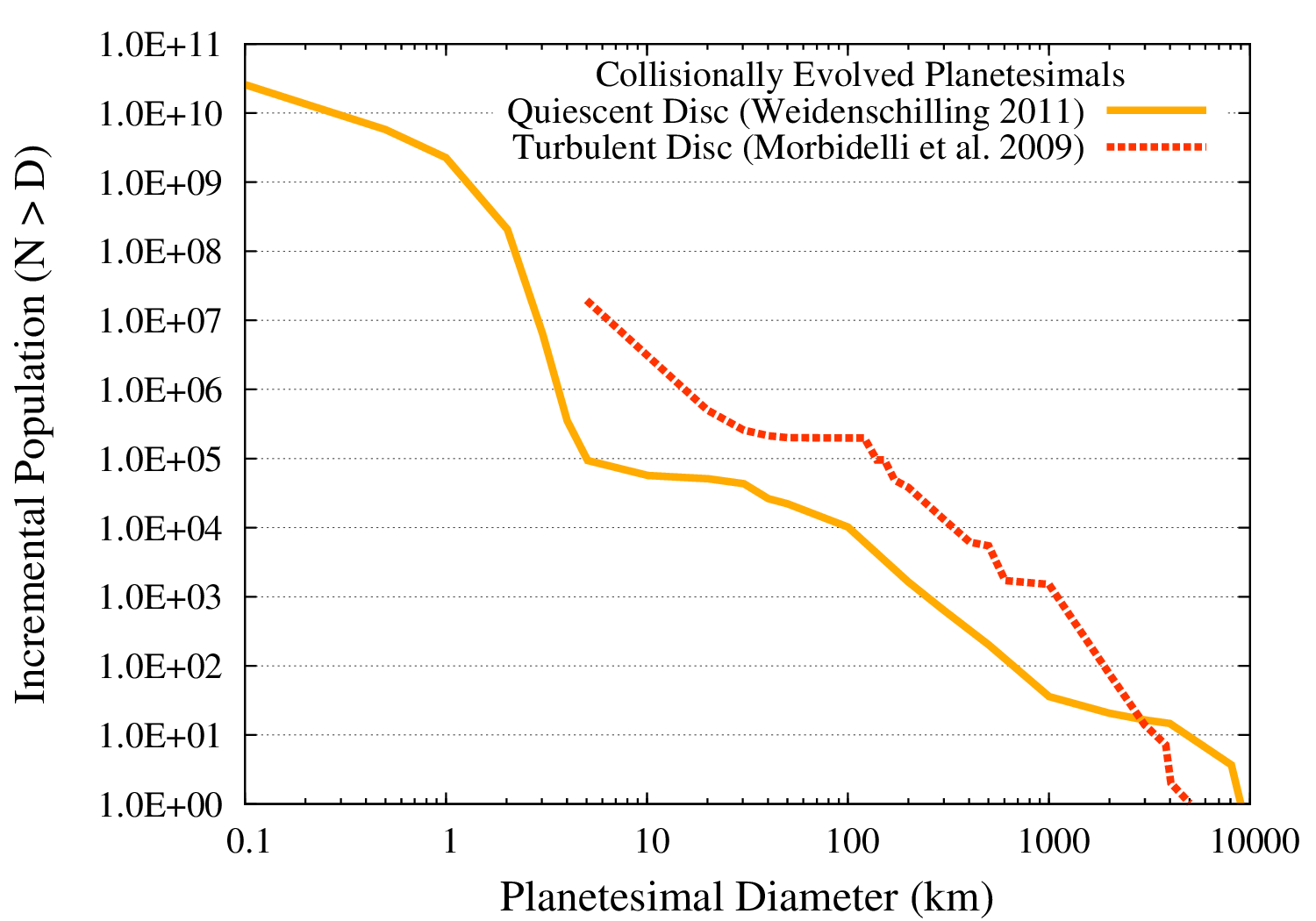}
\caption{Comparison between the two collisionally-evolved SFDs considered in our case study in the orbital region comprised between 2 and 3 au (see Sects. \ref{section-pqce} and \ref{section-ptce} for details).}\label{fig-evolved_sfds}
\end{figure}

}
\subsubsection{Primordial planetesimals formed in a quiescent circumsolar disc}\label{section-pqp}
The first SFD considered was that of a disk of \emph{primordial planetesimals} formed by gravitational instability of the dust in the mid-plane of a \emph{quiescent circumsolar disc}  \citep{safronov1969,goldreich1973,weidenschilling1980,coradini1981}. Following \citet{coradini1981}, the circumsolar disc was assumed to have a density profile $\sigma=\sigma_{0}\left(\frac{r}{1\,AU}\right)^{-n_{s}}$, with $\sigma_{0}=2700$ g cm$^{-2}$ being the {\color{black}gas} surface density at $1$ AU and $n_{s}=1.5$. For this SFD, which we derived from the results of \citet{coradini1981}, the diameters of the planetesimals that could impact Vesta roughly range between 1 and 40 km, with the bulk of the impactors being constituted by planetesimals with diameters of 10-20 km \citep{turrini2014b,turrini2014c}. For more details on the SFD and the associated Monte Carlo method we refer interested readers  to \citet{turrini2014b} and \citet{turrini2014c}.

\subsubsection{Primordial planetesimals formed in a turbulent circumsolar disc}\label{section-ptp}
The second SFD considered was that of \emph{primordial planetesimals} formed by concentration of dust particles in low vorticity regions in a \emph{turbulent circumstellar disc} \citep{cuzzi2008,cuzzi2010}. Following \citet{chambers2010}, the circumstellar disc was assumed to possess a density profile $\sigma=\sigma'_{0}\left(\frac{r}{1\,AU}\right)^{-n'_{s}}$, with $\sigma'_{0}=3500$ g cm$^{-2}$ being the {\color{black}gas} surface density at $1$ AU and $n'_{s}=1$ (see Fig. $14$, gray dot-dashed line, \citealt{chambers2010}). For this SFD, which we derived from the results of \citet{chambers2010}, the diameters of the planetesimals that could impact Vesta roughly range between 20 and 250 km, with the bulk of the impactors being constituted by planetesimals with diameters of 100-200 km \citep{turrini2014b,turrini2014c}. For more details on the SFD and the associated Monte Carlo method we refer interested readers  to \citet{turrini2014b} and \citet{turrini2014c}.

\subsubsection{Collisionally-evolved planetesimals formed in a quiescent circumstellar disc}\label{section-pqce}

The third SFD we considered was associated to \emph{collisionally-evolved planetesimals} formed in a \emph{quiescent circumstellar disc} and was derived from the results of \citet{weidenschilling2011}. In this study we focused on the SFD of the asteroid belt that \citet{weidenschilling2011} referred to as the ``standard case'', i.e. the one produced from a disk initially populated by planetesimals with a diameter of $100$ m (see Fig. $8$, \citealt{weidenschilling2011}). 

The resulting population of planetesimals is dominated \emph{in number} by collisional fragments with km- or sub-km-sized diameters and \emph{in mass} by a few large planetesimals and planetary embryos. In our estimates of the collisional evolution of Vesta we adopted as our lower-end cut-off of the SFD the diameter of 1 km, a choice motivated by the fact that the slope of the SFD causes sub-km planetesimals to cumulatively supply only a fraction of the mass contained in km-sized planetesimals \citep{weidenschilling2011}. 

Because of this cut-off, the bulk of the planetesimals impacting Vesta is in the form of planetesimals with diameters of 1-2 km \citep{turrini2014b,turrini2014c}. Lowering our cut-off to 100 m would increase the mass flux on Vesta only by about 10$\%$ with respect to that provided by km-sized asteroids.

Strictly speaking, the results of \citet{weidenschilling2011} apply only  to the inner Solar System (i.e. $1-4$ au), so in principle they cannot be applied to the outer part of the planetesimal disk (i.e. $4-10$ au) considered by \citet{turrini2011}. However, the results of \citet{weidenschilling2008,weidenschilling2011} suggest that the collisionally-evolved SFD of the planetesimals in our regions of interest does not strongly depend on the radial distance. 

We followed the approach used in \citet{turrini2014c} and adopted a similar SFD for the planetesimals beyond 4 au, scaling it in mass by the ratio between the solid mass comprised between 4 and 10 au and that comprised between 1 and 4 au. For more details on the SFD and the associated Monte Carlo method we refer interested readers  to \citet{turrini2014b} and \citet{turrini2014c}.

\begin{figure}[t]
\centering
\includegraphics[width=\columnwidth]{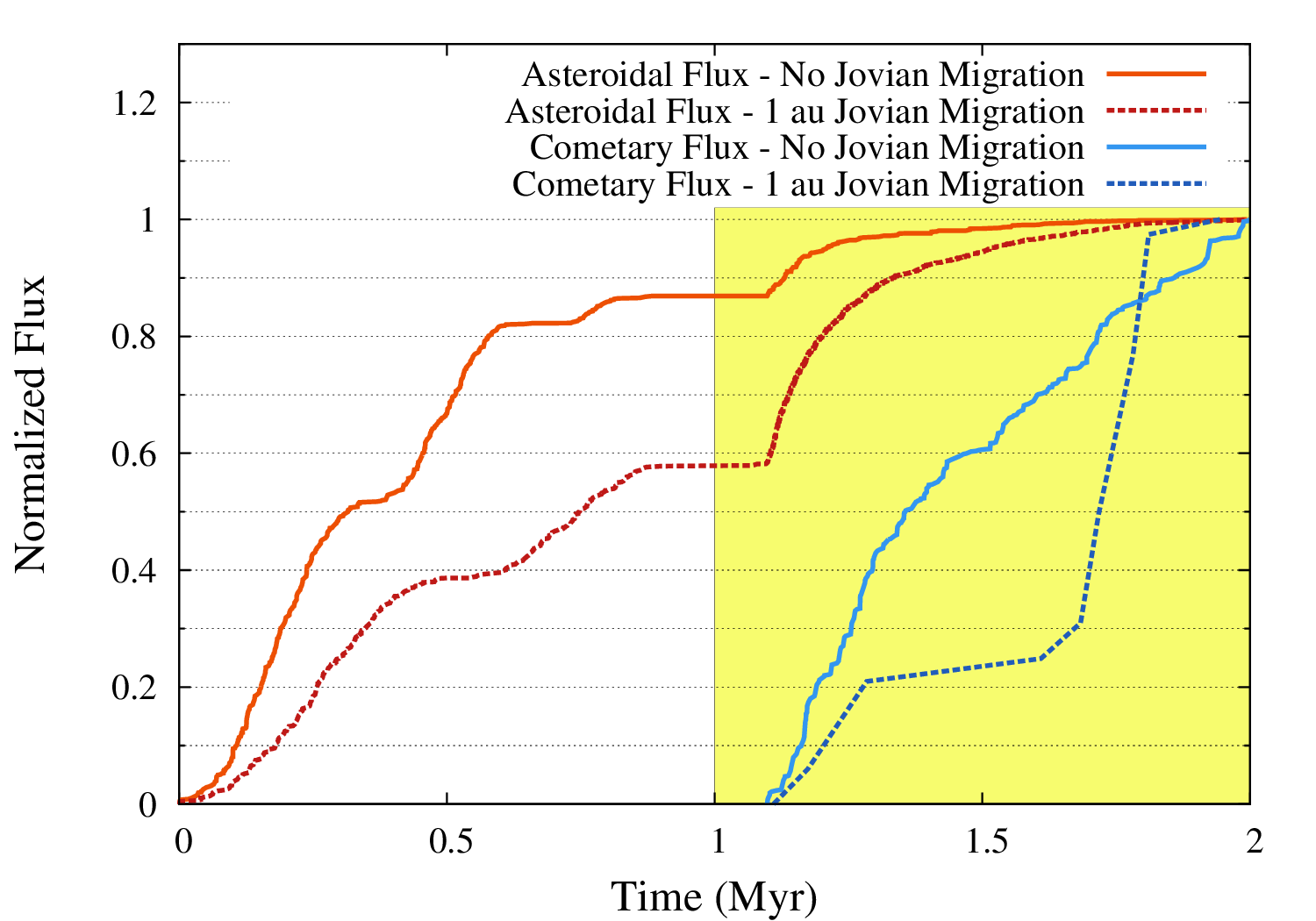}
\caption{Normalized temporal distribution of the fluxes of asteroidal impactors (the orange and red lines) and cometary impactors (the light and dark blue lines) on Vesta in the no migration scenario (the solid lines) and the 1 au migration scenario (the dashed lines) for Jupiter. The highlighted area indicates the temporal interval over which we computed the late accretion and erosion of Vesta's crust, i.e. the Jovian Early Bombardment. Asteroidal impacts before this time were characterized by low velocities ($<$ 1 km/s) and were not considered to account for the clearing effects of Vesta's formation on the orbital region surrounding the asteroid. As can be immediately seen, the Jovian migration enhances the flux of high-velocity ($>$ 1 km/s) asteroidal impactors on Vesta while at the same time decreasing and making more erratic the flux of cometary impactors (see also Fig. \ref{fig-jeb}).}\label{fig-impactors_fluxes}
\end{figure}
\subsubsection{Collisionally-evolved planetesimals formed in a turbulent circumstellar disc}\label{section-ptce}
The fourth and final SFDs we considered was associated to the case of \emph{collisionally-evolved planetesimals} formed in \emph{turbulent circumstellar disc} and was derived from the results of \citet{morbidelli2009}. \citet{morbidelli2009} found that the best match with the present-day SFD of the asteroid belt is obtained for planetesimal sizes initially spanning $100-1000$ km (see Fig. $8$, \citealt{morbidelli2009}), a range consistent with their formation in a turbulent nebula. 

The SFD associated to the best-fit case of \citet{morbidelli2009} shares most of the characteristics of the analogous one derived by \citet{weidenschilling2011}, but shows a larger abundance of planetesimals with diameter comprised between 5 and 20 km (see Fig. $8a$, black solid line, \citealt{morbidelli2009}) than the SFD by \citet{weidenschilling2011}, which is significantly flatter in this size range. 

While the SFD physically extends down to sub-km sizes, we focused our attention on the effects of this overabundance and maintained the lower-end cut-off of the SFD at 5 km in diameter also adopted in \citet{morbidelli2009}. Because of this, the bulk of the planetesimals impacting Vesta is in the form of planetesimals with diameters of 5-10 km \citep{turrini2014b,turrini2014c}.

As in the case of the SFD by \citet{weidenschilling2011} discussed in Sect. \ref{section-pqce}, we extended the SFD of \citet{morbidelli2009} beyond 4 au by scaling the number of planetesimals by a factor equal to the mass ratio of the solid material contained between 4 and 10 au to that of the one contained between 1 and 4 au. For more details on the SFD and the associated Monte Carlo method we refer interested readers  to \citet{turrini2014b} and \citet{turrini2014c}.

\begin{figure}[t]
\centering
\includegraphics[width=\columnwidth]{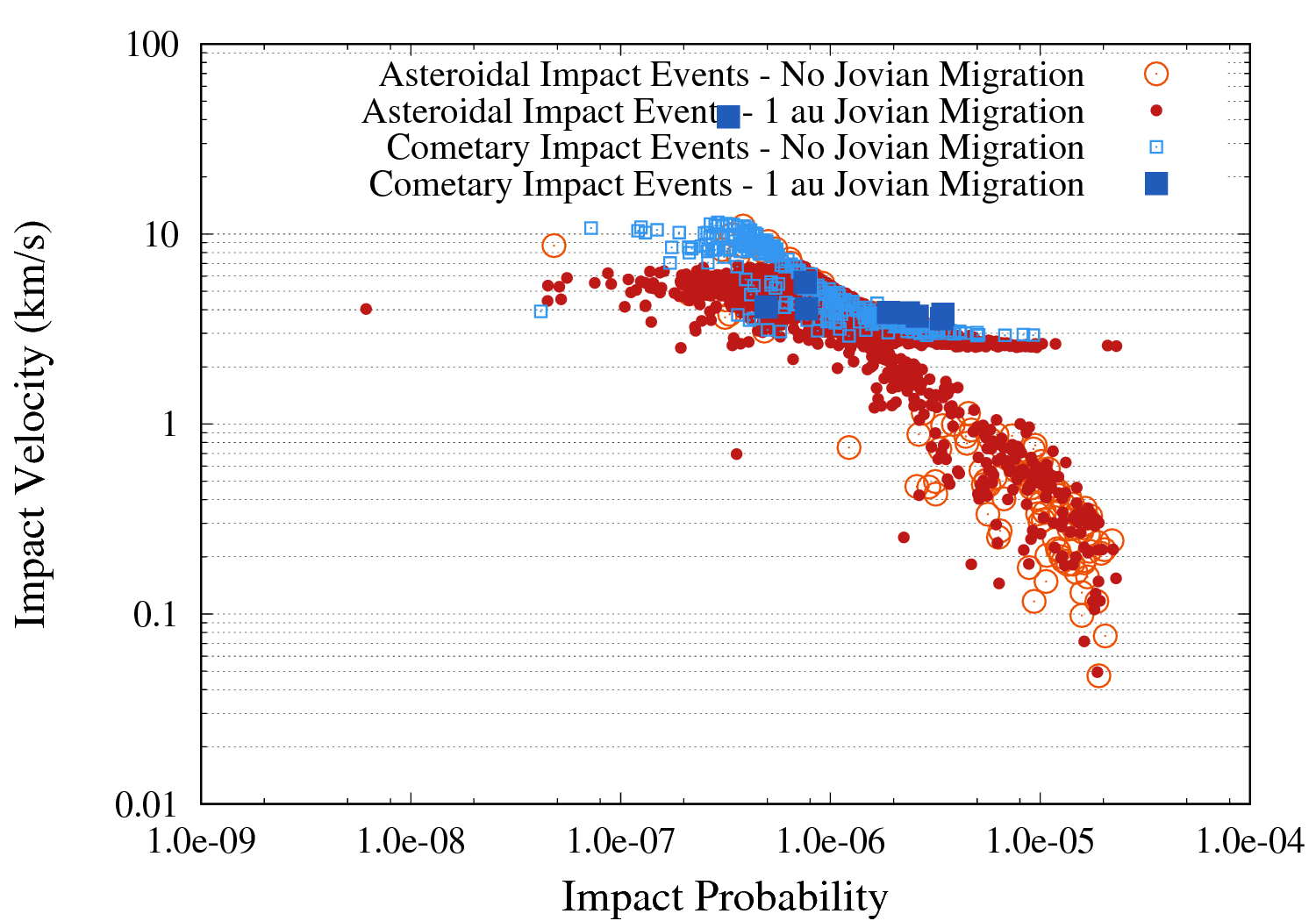}
\caption{Distribution of the impact probabilites and impact velocities of the asteroidal and cometary impactors in the scenario of no migration of Jupiter and in the 1 au migration scenario for the giant planet in the simulations from \citet{turrini2011}. Note that the impact probabilities reported here refer to the individual impact events and are not impact probabilities averaged over the whole populations of impactors as in classical collisional algorithms (see e.g. \citealt{obrien2011} and references therein).}\label{fig-events_comparison}
\end{figure}

\begin{figure*}[t]
\centering
\includegraphics[width=\textwidth]{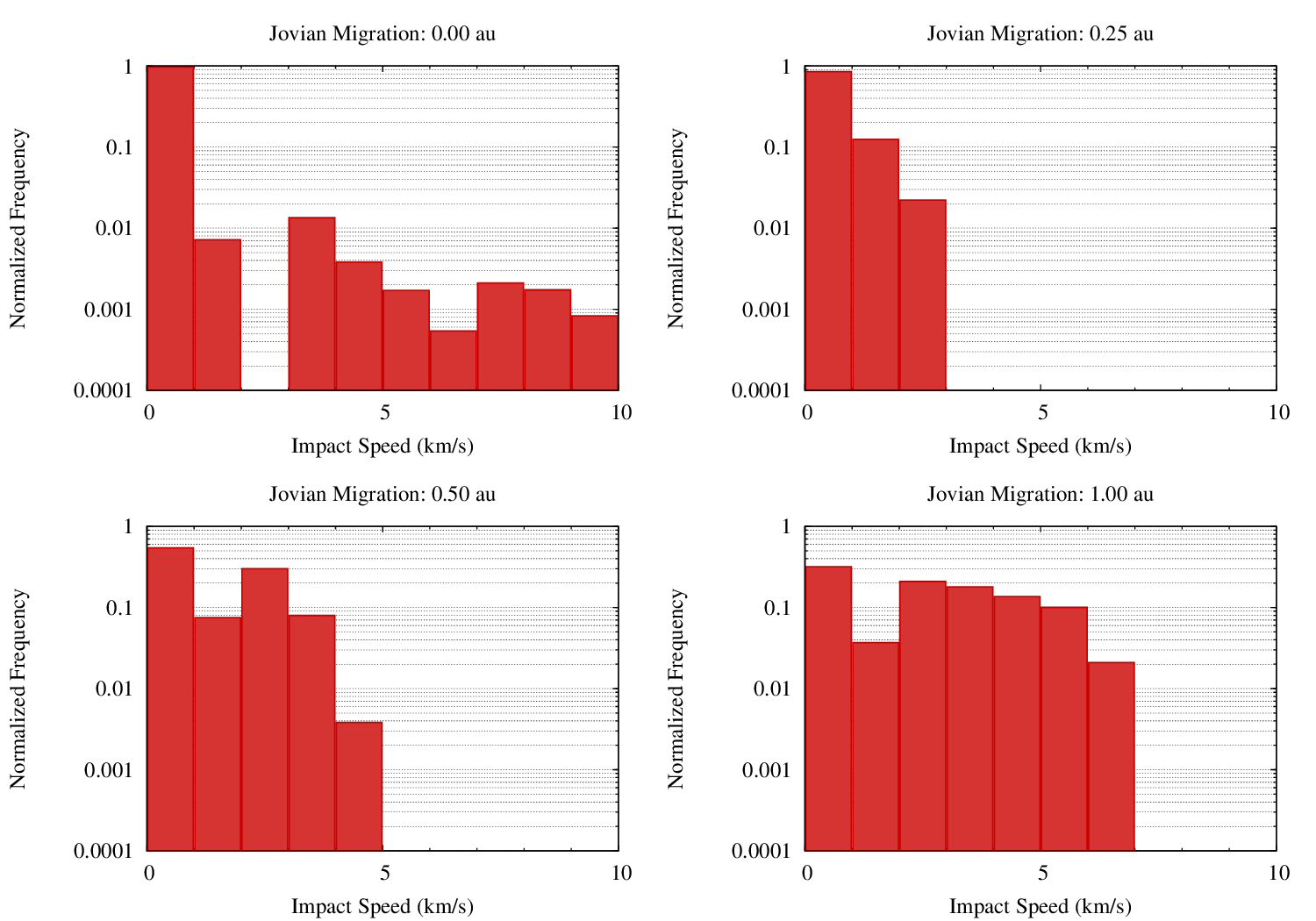}
\caption{Normalized distribution of the impact velocities of the asteroidal impactors (i.e. the impactors originating between 1 and 4 au in the simulations of \citealt{turrini2011}) on Vesta in the four migration scenarios considered in our case study (see \citealt{turrini2011} and \citealt{turrini2014b} for more details).}\label{fig-asteroidal_velocities}
\end{figure*}

\begin{figure*}[t]
\centering
\includegraphics[width=\textwidth]{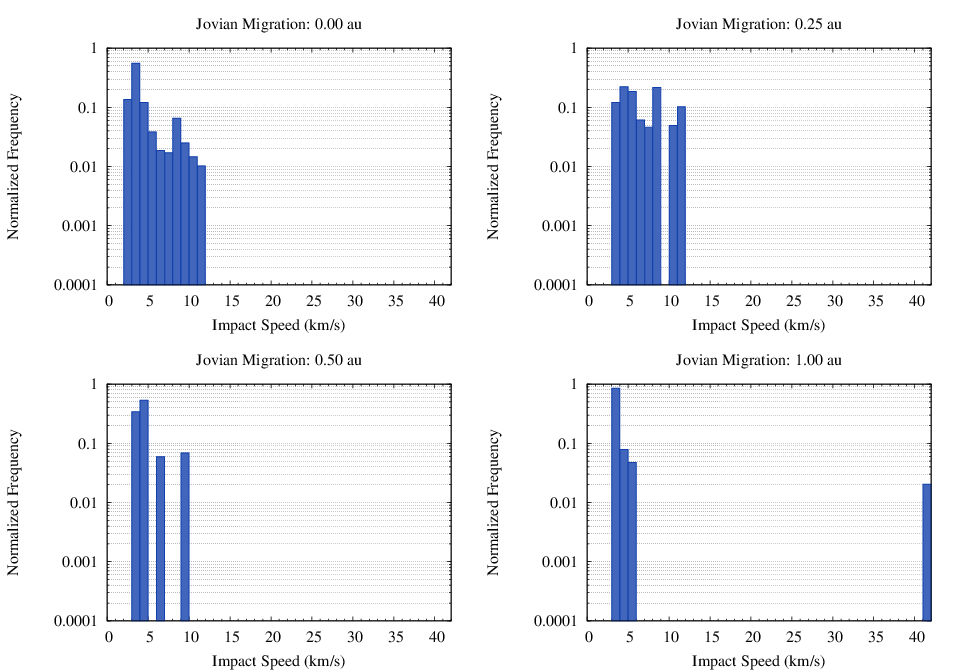}
\caption{Normalized distribution of the impact velocities of the cometary impactors (i.e. the impactors originating between 4 and 10 au in the simulatios of \citealt{turrini2011}) on Vesta in the four migration scenarios considered in our case study (see \citealt{turrini2011} and \citealt{turrini2014c} for more details).}\label{fig-cometary_velocities}
\end{figure*}

\subsection{Modelling Vesta's collisional history}\label{section-collisions}

\citet{turrini2011} estimated the impact probabilities and the associated impact velocities between the massless particles and Vesta using a statistical approach based on solving the ray--torus intersection problem between the {\color{black}instantaneous} orbital torus of Vesta and the linearized path of {\color{black}the massless particle\footnote{{\color{black}Note that the path of the massless particle is linearized only for the computation of its impact probability with Vesta, not for that of the dynamical evolution of the particle.}} across the time step when the particle crosses Vesta's orbital region} {\color{black}(see \citealt{turrini2011} for more details on the method)}. This method is conceptually similar to the analytical method of \citet{opik1976} but requires only to average over the mean anomaly of the target body's orbit {\color{black}instead of averaging on anomaly, longitude of nodes and argument of pericenter of both target and impacting bodies. 

In evaluating the collisional history of Vesta we focused on the massless particles impacting Vesta from the moment Jupiter's core started accreting its gaseous envelope (i.e. the second 1 Myr in the simulations by \citealt{turrini2011}, see the highlighted area in Fig. \ref{fig-impactors_fluxes}). This conservative choice is motivated by the need to correct for the fact that the early flux of impactors on Vesta in the simulations is dominated by the impacts of those rocky planetesimals orbiting nearby the asteroid that should have been removed during Vesta's formation. 

Fig. \ref{fig-events_comparison} shows an example of the distributions of impact probabilities and impact velocities for both asteroidal and cometary impactors recorded in the simulations by \citet{turrini2011} in the scenarios of no migration and 1 au migration of Jupiter. Note that the impact probabilities reported in Fig. \ref{fig-events_comparison} refer to the individual impact events recorded in the simulations and are not impact probabilities averaged over the whole populations of impactors as in classical collisional algorithms (see e.g. \citealt{obrien2011} and references therein). Figs. \ref{fig-asteroidal_velocities} and \ref{fig-cometary_velocities} show respectively the  distributions normalized over the impact probabilities of the asteroidal and cometary impact velocities in the four migration scenarios considered in this study (see also \citealt{turrini2011}, \citealt{turrini2014b} and \citealt{turrini2014c} for a more detailed discussion of the distribution of the impact velocities and their causes.} Interested readers are referred to \citet{turrini2011} {\color{black}and \citet{turrini2012} for details on the algorithm.}

The impact probabilities provided by the simulations were converted into fluxes of impactors using the SFDs described in Sect. \ref{section-planetesimals}. Following the procedure described in \citet{turrini2014b} and \citet{turrini2014c}, for each SFD we run a set of $10^{4}$ Monte Carlo simulations. In each run a new mass value was extracted for each impact event recorded in \citet{turrini2011} and, {\color{black}since each massless particle causing an impact event represents a swarm of real planetesimals, we used the SFD and the impact probability of the impact event to estimate the associated flux of impactors. Combining the information provided by the mass and flux of impactors associated to the impact event with its estimated impact velocity, the eroded mass $m_{e}$ and the accreted mass $m_{a}$ were computed (see Sect. \ref{section-impacts} for details on the method).} 

We averaged over each set of $10^{4}$ Monte Carlo simulations to estimate the total mass loss and accretion experienced by Vesta for each specific SFD {\color{black}and the associated standard deviations}. {\color{black}If, after averaging, the total flux of impactors associated to one of the SFDs amounted to less than one real impact, we set the total mass loss and accretion values to zero for that SFD.

{\color{black}\subsection{Modelling the effects of impacts on Vesta}\label{section-impacts}}


To estimate the effects of impacts in terms of both mass loss and mass accretion, we took advantage of the results of \citet{benz1999} (see Sect. \ref{section-massloss} for details) and \citet{turrini2016} (see Sect. \ref{section-massgain} for details). In parallel, we performed 3D numerical simulations of impacts of projectiles onto Vesta using a modified version \citep{svetsov2011,turrini2014c,svetsov2015} of the numerical hydrodynamic method SOVA (\citealt{shuvalov1999}; SOVA is an acronym for Solid-Vapour-Air, as the code is designed for simulations of multi-material, multi-phase flows) that includes the effects of dry friction \citep{dienes1970}. 

Dry friction depends on a dimensionless coefficient of friction for which we adopted a value of 0.7, typical for rocks and sand \citep{turrini2014c,turrini2016}. The behaviour and properties of target and projectiles were determined, as in \citet{turrini2014c} and \citet{turrini2016}, through the ANEOS equations of state \citep{thompson1972} using input data (i.e., about 35 variables describing properties of a given material) from \citet{pierazzo1997} and Tillotson's equation of state for Vesta's iron core \citep{tillotson1962}. 

In the simulations performed with SOVA, Vesta was modelled as a three-layered sphere with radius of $260$ km, possessing an iron core with a radius of $110$ km \citep{russell2012,russell2013,ermakov2014} and a crust made of granite with a thickness of $23$ km \citep{consolmagno2015}, separated by a mantle composed of dunite. The mass of Vesta was set equal to its present value, $2.59\times10^{23}$ g \citep{russell2012}. {\color{black}While Vesta was in a partially molten state at the time of the Jovian Early Bombardment, the approximation we adopted is justified by the following reasons.

First, thermal and geophysical models and meteoritic data all suggest that Vesta's basaltic crust was formed over a series of magmatic effusive events through a solid conductive lid. Second, previous studies indicates that Vesta's mass loss due to cratering erosion was mainly a surface process \citep{turrini2014b,turrini2014c}, hence mainly affecting this solid conductive lid. Third, mass loss occurs mainly from the central regions of the crater where the material strength is generally unimportant \citep{holsapple2007}, since the stresses during the impacts exceed the strength of the excavated material acquiring velocities greater than the escape velocity of the asteroid. This approximation, however, is more realistic for impactors not exceeding in size the thickness of Vesta's conductive lid (i.e. a few km) than for larger impactors.
} 

As in \citet{turrini2016}, the numerical grid consisted of $250\times100\times225$ cells over azimuth, polar angle and radial distance respectively, and we assumed bilateral symmetry to model only the half-space in the zenith direction. Cell sizes were $1/40$ of the projectile's diameter around the impact point and increased to the antipodal point and to the radial boundaries located at distances of about $10$ vestan radii. In all impact simulations, the impact velocity vector lied in the reference plane that passed through the origin of the coordinates and was orthogonal to the zenith.

All simulated impacts were assumed to occur at the average impact angle of 45$^{\circ}$ \citep{melosh1989}, while impact velocities varied between 1 and 12 km/s based on the results of the simulations performed by \citet{turrini2011} (see {\color{black}Figs. \ref{fig-asteroidal_velocities} and \ref{fig-cometary_velocities}} and \citealt{turrini2014b,turrini2014c} for more details on the distribution of the impact velocities in the different migration scenarios).

We performed simulations of cometary impactors composed by a homogeneous mixture of rocks and ices (see \citealt{svetsov2015}, Fig. 5). Among the materials supplied by the ANEOS equations of state \citep{thompson1972}, we adopted water as our template for the icy component and granite as our template for the rocky one. The simulations described in \citet{turrini2016} provided us with analogous results for asteroidal rocky impactors. 

Among the different kinds of rocky impactors (granite impactors, dunite impactors and differentiated impactors
) simulated by \citet{turrini2016} we adopted their results for granite impactors as our template for asteroidal impactors. The comparison between the results of impact experiments \citep{holsapple1993,holsapple2007,daly2016} and those of  SOVA's simulations reveals that they agree within a factor of two \citep{svetsov2011,turrini2016}.  

\subsubsection{Mass loss associated to the impact events}\label{section-massloss}

Following \citet{turrini2014b} and \citet{turrini2014c}, we defined three classes of impact events based on their normalized specific energy $Q_{D}/Q^{*}_{D}$, where $Q^{*}_{D}$ is the catastrophic disruption threshold of Vesta. Impacts with $Q_{D}/Q^{*}_{D} < 0.1$ were classified as \emph{low-energy impacts}. Impacts with $0.1 \leq Q_{D}/Q^{*}_{D} < 1$ were classified as \emph{high-energy impacts}. Impacts with $Q_{D}/Q^{*}_{D} \geq 1$ were classified as \emph{catastrophic impacts}.

\begin{figure}
\centering
\includegraphics[width=\columnwidth]{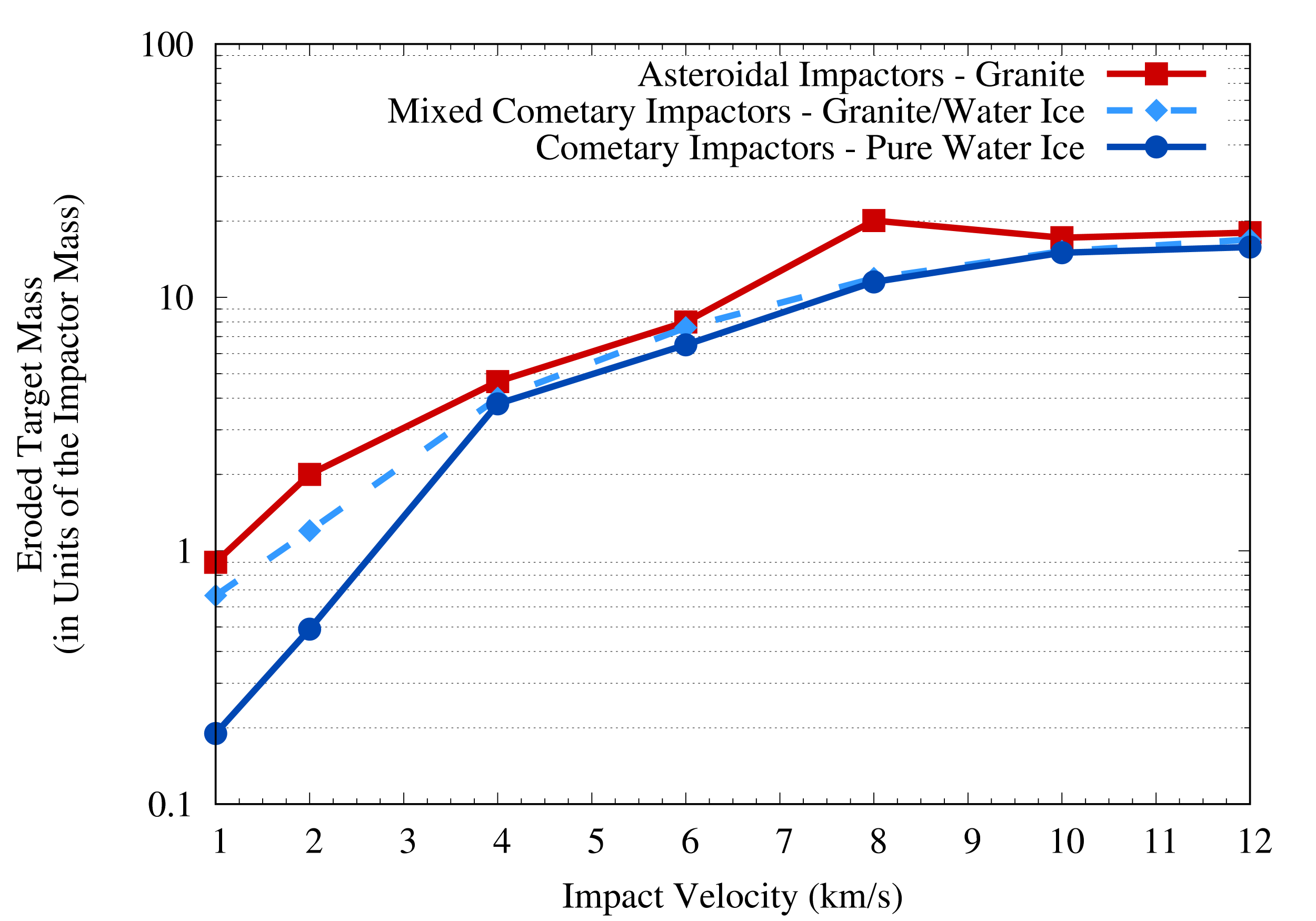}
\caption{Fraction of the mass of the target body Vesta that is eroded and lost due to the impact, in units of the mass of the projectile. The different curves show the results from the simulations of \citet{turrini2016} for asteroidal impactors made of granite (red solid line with filled squares), the simulations performed in this work for mixed granite-water ice cometary impactors (light blue dashed lines with filled diamonds), and, for comparisons, the results of the simulations of \citet{turrini2014c} for cometary impactors made of pure water ice (blue solid line with filled circles).}\label{fig-erosion}
\end{figure}

The quantity $Q^{*}_{D}$ was computed using Eq. $6$ from \citet{benz1999} with the associated coefficients for basaltic targets (see Table $3$, \citealt{benz1999}). Following \citet{turrini2014b} and \citet{turrini2014c}, we used the coefficients of the case $v_{i}=5\,km\,s^{-1}$ for impacts with velocity greater or equal than $5\,km\,s^{-1}$, and those of the case $v_{i}=3\,km\,s^{-1}$ for all the other impacts.

We computed the mass loss associated to low-energy impacts using the results of the impact simulations with SOVA performed in the framework of this study and those performed by \citet{turrini2016}. The results of the simulations are shown in Fig. \ref{fig-erosion}, where the mass loss  as a function of the impact velocity is expressed in units of the mass of the impacting body. For comparison, in Fig. \ref{fig-erosion} we also plotted the results of the simulations by \citet{turrini2014c} for cometary impactors composed of pure water ice.

For high-energy impacts we used instead Eq. $8$ from \citet{benz1999} expressed in terms of the eroded mass: 
\begin{equation}\label{eqn-erosion_benz}
\frac{m_{e}}{m_{t}}=0.5+s\left(\frac{Q_{D}}{Q^{*}_{D}}-1.0\right)
\end{equation}
where $s=0.5$ for $v_{i}<5\,km\,s^{-1}$ and $s=0.35$ for $v_{i} \geq 5\,km\,s^{-1}$.
{\color{black} To avoid overestimating the contribution of high-energy impacts to Vesta's crustal erosion, the effects of those high-energy impact events that, after renormalizing to the appropriate SFD, were associated to less than one real impact were not considered in estimating Vesta's crustal erosion}.

The effects of catastrophic impacts were not accounted for in the estimates of the eroded mass: their cumulative number was used only to assess the probability of Vesta surviving its primordial collisional evolution without being shattered (see also \citealt{turrini2014b,turrini2014c} for a discussion).

\subsubsection{Mass gain associated to the impact events}\label{section-massgain}

\begin{figure}
\centering
\includegraphics[width=\columnwidth]{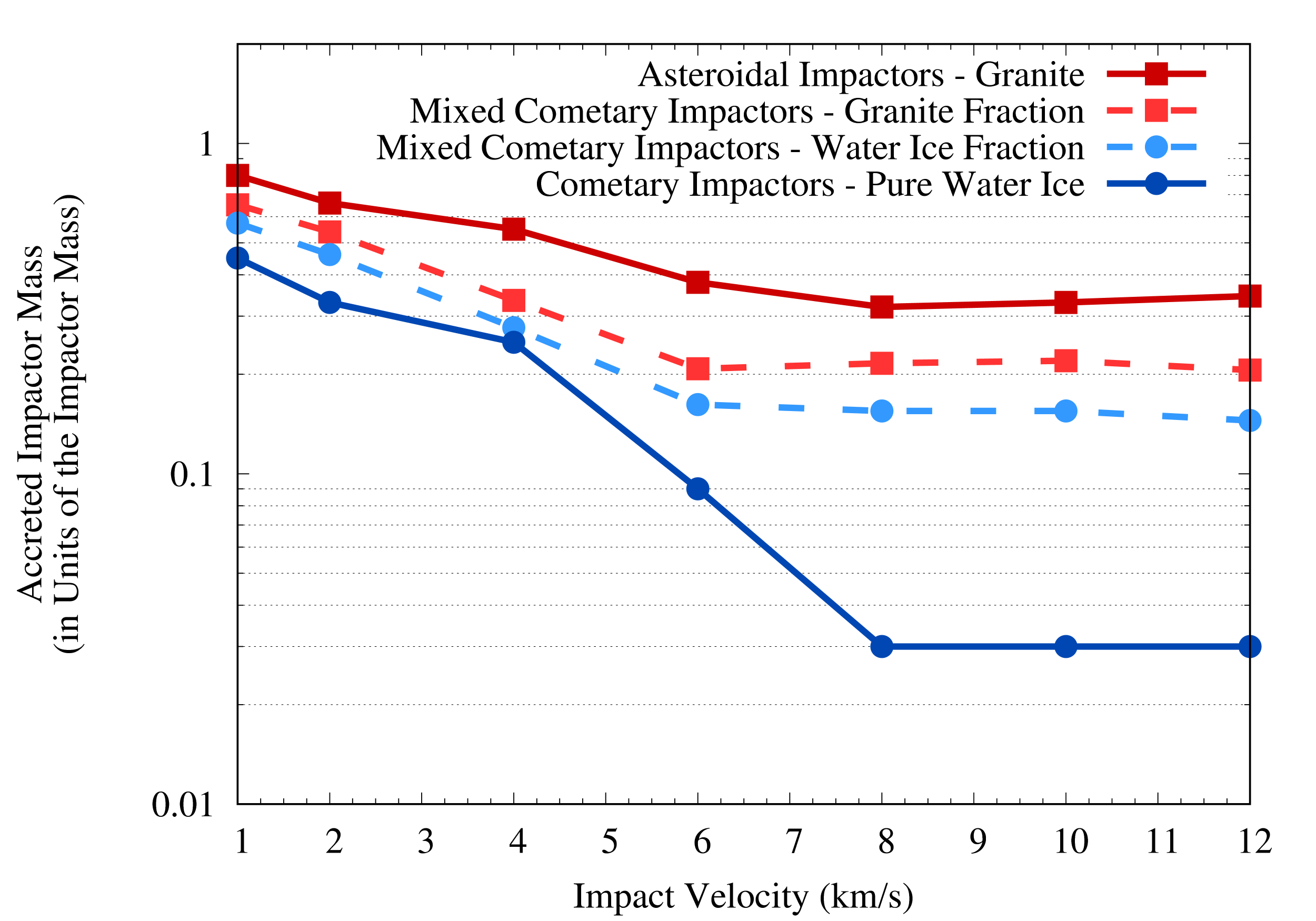}
\caption{Fraction of the mass of the projectile that survives the impact and is accreted by Vesta, in units of the mass of the projectile. The different curves show the results from the simulations of \citet{turrini2016} for asteroidal impactors made of granite (red solid line with filled squares), the simulations performed in this work for mixed granite-water ice cometary impactors (red dashed lines with filled squares for the rocky component and blue dashed lines with filles circles for the icy component), and, for comparisons, the results of the simulations of \citet{turrini2014c} for cometary impactors made of pure water ice (blue solid line with filled circles).}\label{fig-accretion}
\end{figure}

To assess the mass accretion experienced by primordial Vesta we again took advantage of the results of the impact simulations with SOVA performed in the framework of this study and those performed by \citet{turrini2016}. The results of the simulations are shown in Fig. \ref{fig-accretion}, where the accreted mass as a function of the impact velocity is expressed in units of the mass of the impacting body. For comparison, in Fig. \ref{fig-accretion} we also plotted the results of the simulations by \citet{turrini2014c} for cometary impactors composed of pure water ice. 

The results of the simulations in \citet{turrini2016} indicated that the composition and the diameter of rocky impactors do not change the results of the simulations as much as the impact velocity (i.e. the effects of the former parameters are limited to about $5-10\%$, see \citealt{turrini2016} for a discussion). Both low-energy and high-energy ones contributed mass to Vesta according to the results shown in Fig. \ref{fig-accretion}, while catastrophic impact did not contribute mass to Vesta. {\color{black} For consistency with the procedure adopted in estimating the mass loss caused by high-energy impacts, the contribution of those high-energy impact events that, after renormalizing to the appropriate SFD, were associated to less than one real impact was not considered in estimating Vesta's late accretion}.

\section{Results}\label{section-results}

{\color{black}In the following we present the late accretion and erosion experienced by Vesta's crust across Jupiter's formation and migration, as depicted by our results taken at face value. For each of the four SFDs we considered we will show the average mass loss, mass accretion and water accretion produced by Vesta's early collisional evolution. We will first discuss the separate contributions of asteroidal and cometary impactors, which are defined as those planetesimals originating within and beyond 4 au respectively, and then their cumulative effects on Vesta. When considering the cumulative collisional history of the asteroid, we will discuss how it affects both a primordial Vesta similar in mass to the present one (\emph{``intact and pristine Vesta''} scenario) and a Vesta two to three times larger (\emph{``altered Vesta''} scenario).

For each of the average quantities we computed, we will also show the associated standard deviations as a measure of the variability of our results. The two main factors affecting the magnitude of the standard deviations are the total flux of impactors and the variability of the number of the largest impactors (see e.g. \citealt{turrini2014d,turrini2016}). As such, the largest standard deviations will be associated to the populations of cometary impactors (more affected by the effects of small-number statistics due to their lower fluxes) and to the population of collisionally-evolved impactors formed in turbulent discs (due to the effects of small-number statistics on the flux of large impactors).}
}

\subsection{Mass loss and crustal erosion}

\begin{figure*}[t]
\centering
\includegraphics[width=\textwidth]{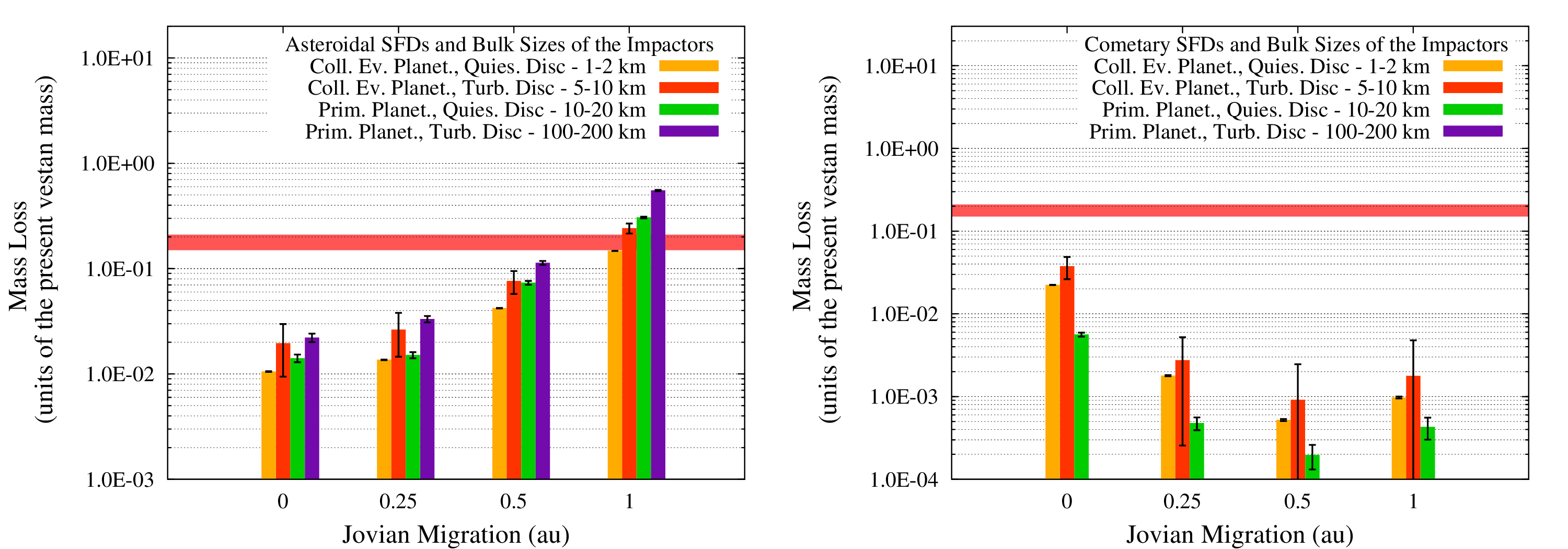}
\caption{Mass loss experienced by a primordial Vesta with mass similar to that of the present Vesta due to (\emph{left}) asteroidal impactors and (\emph{right}) cometary impactors during Jupiter's mass growth in the different migration scenarios and for the different SFDs considered. For each SFD we report the characteristic diameter of the planetesimals producing the bulk of the impact flux as computed with our Monte Carlo methods. The horizontal regions highlighted in {\color{black}red} mark the range of values of Vesta's crustal mass fraction {\color{black} and represent our upper boundary to Vesta's mass loss (see Sect. \ref{section-constraints} and \citealt{consolmagno2015}). Note that, given that the temporal interval considered in this proof-of-concept study is smaller than the timespan over which Vesta's crust can be eroded, only those scenarios producing mass losses \emph{below} the red regions should be considered compatible with present-day Vesta.}}\label{fig-massloss}
\end{figure*}

The first step of our analysis focused on the mass loss suffered by primordial Vesta {\color{black}in the classical ``intact and pristine Vesta'' scenario, where the asteroid always possessed a mass similar to its present one}. The mass loss caused by asteroidal and cometary impactors individually is shown in Fig. \ref{fig-massloss} and is dominated by the effects of low-energy impacts (see also \citealt{turrini2014b,turrini2014c}). Catastrophic impacts have a limited probability to occur (generally less than $0.1\%$ and never above $1\%$). 

High-energy impacts are comparatively more probable in the case of the SFDs associated with a turbulent circumsolar disc. Also in those cases, however, the chances of high-energy impacts occurring never exceed $20-30\%$. The only notable exception is the case of primordial planetesimals formed in a turbulent circumsolar disc \citep{chambers2010} when Jupiter migrates by 1 au, where Vesta could experience two high-energy impacts (responsible for about $60\%$ of the total mass loss associated to this SFD in this migration scenario).
 
The mass loss experienced by Vesta due to asteroidal impactors (Fig. \ref{fig-massloss}, left panel) is limited in the cases of no migration or 0.25 au of migration of Jupiter but experiences a rapid growth once Jupiter's migration reaches and exceeds 0.5 au. The initial limited mass loss, of the order of $\sim1\%$, is mainly due to impactors excited by the 3:1 resonance with Jupiter. When Jupiter's migration reaches 0.5 au a second family of higher-velocity impactors excited by the 2:1 resonance with Jupiter appears (see Fig. \ref{fig-jeb} and \citealt{turrini2011}). This second family causes the mass loss experienced by Vesta to grow by about an order of magnitude.

The mass loss associated to cometary impactors shows an opposite trend, being significant only when Jupiter does not experience migration and dropping by more than one order of magnitude in those scenarios where the giant planet migrates (see Fig. \ref{fig-massloss}, right panel). This is due to the fact that the migration of the giant planet favours the trapping of more and more planetesimals in the sweeping resonances at the outer boundaries of the asteroid belt, reducing  Jupiter's efficiency in scattering cometary planetesimals in the orbital region of Vesta (see Fig. \ref{fig-jeb} and \citealt{turrini2011}).

\begin{figure*}
\centering
\includegraphics[width=\textwidth]{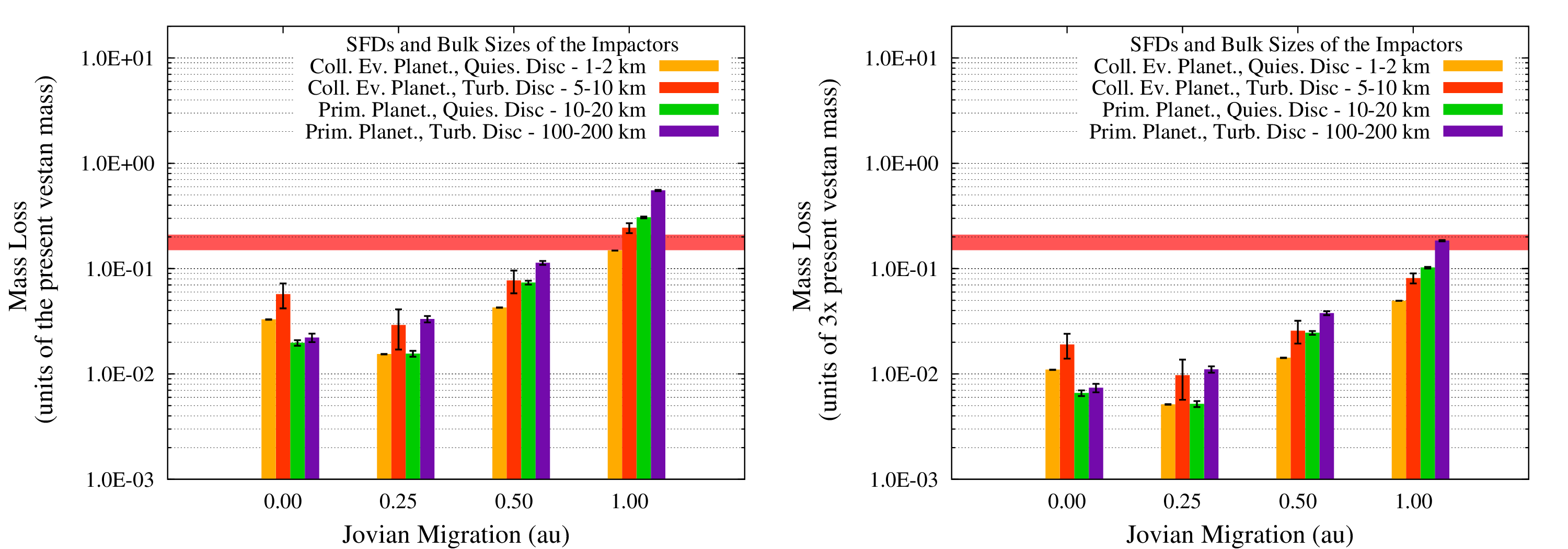}
\caption{Total mass loss experienced by (\emph{left}) a primordial Vesta with the same mass as present Vesta and (\emph{right}) a primordial Vesta three times as massive during Jupiter's mass growth in the different migration scenarios and for the different SFDs considered. For each SFD we report the characteristic diameter of the planetesimals producing the bulk of the impact flux as computed with our Monte Carlo methods. The horizontal regions highlighted in {\color{black}red} mark the range of values of Vesta's crustal mass fraction {\color{black} and represent our upper boundary to Vesta's mass loss (see Sect. \ref{section-constraints} and \citealt{consolmagno2015}). Note that, given that the temporal interval considered in this proof-of-concept study is smaller than the timespan over which Vesta's crust can be eroded, only those scenarios producing mass losses \emph{below} the red regions should be considered compatible with present-day Vesta.}}\label{fig-massloss-total}
\end{figure*}

The total mass loss experienced by Vesta in the different scenarios is shown in Fig. \ref{fig-massloss-total}. As can be immediately seen, the order of magnitude of the mass loss experienced by Vesta is mainly a function of Jovian migration. The actual SFD of the impacting planetesimals appears to affect the result, within a given migration scenario, to roughly a factor of three. Fig. \ref{fig-massloss-total} reveals that the most favourable cases in terms of experienced mass loss and preservation of the vestan crust are that of a Jovian displacement of 0.25 au and that of no migration of the giant planet.

The cases of a Jovian migration of 0.5 and 1 au appear less favourables and, for a primordial Vesta characterized by a mass similar to its present one, they appear inconsistent with the survival of Vesta's crust (especially once the excavation caused by the two vestan South polar impact basins is taken into account). The case of a Jovian migration of 1 au, in particular, is associated to a mass loss of the same order as the expected mass of the vestan crust.

{\color{black}We then moved to investigate how the picture depicted by these results would change in the ``altered Vesta'' scenario, where primordial Vesta is hypothesized to have been more massive than its present counterpart \citep{consolmagno2015}.} For a primordial Vesta twice as massive as present Vesta, the radius of the asteroid would be larger by about $25\%$ than the present one and the escape velocity would increase by about $100$ m/s, i.e about $30\%$. The increase in the escape velocity would lower the average efficiency of impacts in causing mass loss by about $30\%$ (see Eq. 3 in \citealt{svetsov2011}). As the flux of impactors on Vesta is directly proportional to the radius of the asteroid, the increase in the radius would translate into a similar increase in the flux of impactors (see \citealt{turrini2011} for details). The new flux almost compensates for the decrease in the erosion efficiency of the impacts, so that the overall erosion decreases by about 10$\%$.

Because of this, the values plotted in Figs. \ref{fig-massloss} and \ref{fig-massloss-total} would scale down by slightly more than the mass ratio between the primordial Vesta and the present one. For a primordial Vesta twice as massive as the present one, these values would decrease by a factor of two. The only scenario incompatible with the constraint on Vesta's mass loss would become that of a Jovian migration of 1 au (either due to the mass loss per se or to its combination with the {\color{black}later} excavation caused by the South polar basins). 

A larger primordial mass of Vesta would proportionally decrease the mass lost by the asteroid due to collisions. For a primordial Vesta three times as massive as the present one (see Fig. \ref{fig-massloss-total}), the only cases that would be rejected by the constraint on the crustal survival would be those where Jupiter migrated by 1 au and the flux of impactors on Vesta was dominated by planetesimals with diameters larger than 10 km, as in the SFDs by \citet{coradini1981} and \citet{chambers2010}. 

\subsection{Mass accretion and water delivery}

As discussed in Sects. \ref{section-constraints} and \ref{section-model}, the impacts on Vesta would also cause the asteroid to experience a phase of late accretion. The second step of our analysis was to quantify how much water would be delivered to Vesta by the two potential sources we considered, volatile-rich asteroids and ice-rich comets (see Sects. \ref{section-constraints} and \ref{section-model}), and compare the estimated amounts with the upper bound set by the presence of apatites in basaltic eucrites.  {\color{black}Again, we started with the classical ``intact and pristine Vesta'' scenario, where the asteroid always possessed a mass similar to its present one}.

\begin{figure*}[t]
\centering
\includegraphics[width=\textwidth]{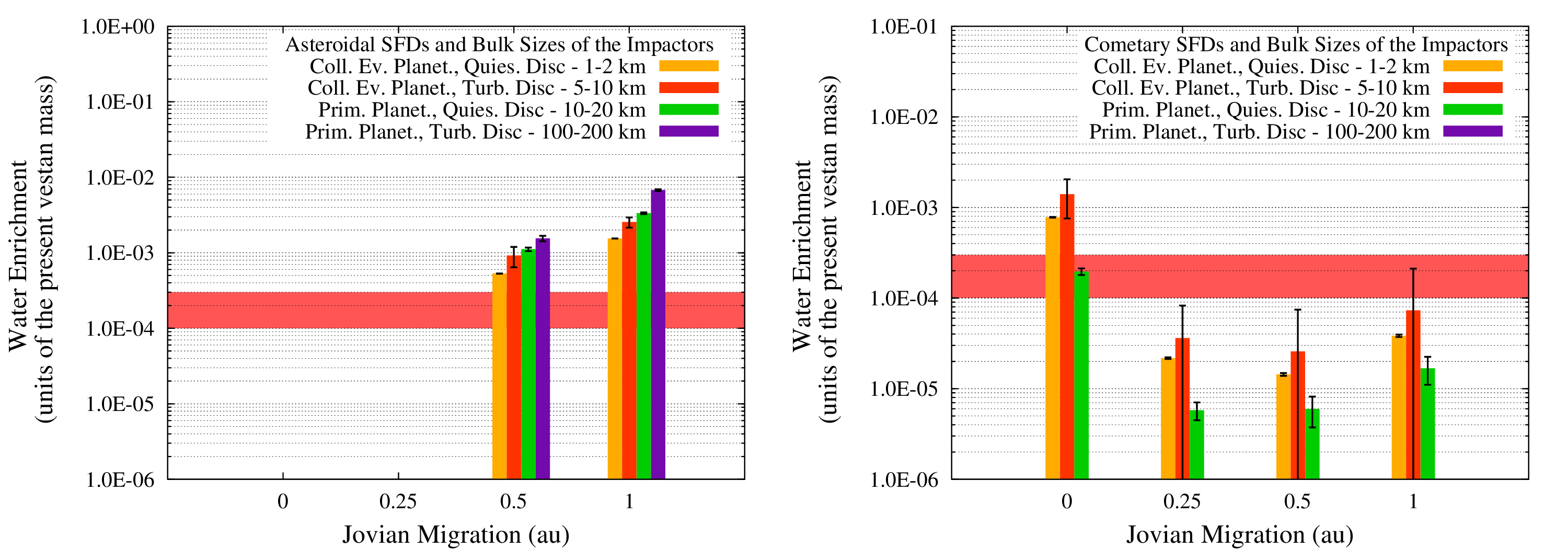}
\caption{Water accretion experienced by a primordial Vesta with mass similar to that of the present Vesta due to (\emph{left}) asteroidal impactors and (\emph{right}) cometary impactors during Jupiter's mass growth in the different migration scenarios and for the different SFDs considered. For each SFD we report the characteristic diameter of the planetesimals producing the bulk of the impact flux as computed with our Monte Carlo methods. The horizontal regions highlighted in {\color{black}red} mark the range of values of Vesta's water enrichment {\color{black} and represent our upper boundary to Vesta's water accretion (see Sect. \ref{section-constraints} and \citealt{stephant2016a,stephant2016b,sarafian2017a,sarafian2017b}). Note that, given that the temporal interval considered in this proof-of-concept study is smaller than the timespan over which Vesta's crust can be enriched in water, only those scenarios producing water enrichments \emph{below} the red regions should be considered compatible with present-day Vesta.}}\label{fig-water}
\end{figure*}

The individual contributions of asteroids and comets are shown in Fig. \ref{fig-water}. Asteroidal impactors (Fig. \ref{fig-water}, left panel) deliver water to Vesta only when the Jovian migration reaches or exceeds 0.5 au, as the dynamical excitation of the population of planetesimals affected by the sweeping 2:1 resonance with Jupiter allows them to reach the orbital region of Vesta and deliver water to the asteroid (see Fig. \ref{fig-jeb} and \citealt{turrini2011}).

The case of cometary impactors (Fig. \ref{fig-water}, right panel) is opposite to that of the asteroidal ones, as they deliver significant amounts of water to Vesta only when Jupiter does not migrate. If the giant planet migrates, the amount of water accreted by Vesta drops by more than one order of magnitude, showing however a slowly increasing trend with increasing displacements of Jupiter. The SFD associated to primordial planetesimals formed in a turbulent circumstellar disc (see Sect. \ref{section-ptp}) does not appear in the right panel of Fig. \ref{fig-water} as its total flux amounts to less than one impact event.

\begin{figure*}[t]
\centering
\includegraphics[width=\textwidth]{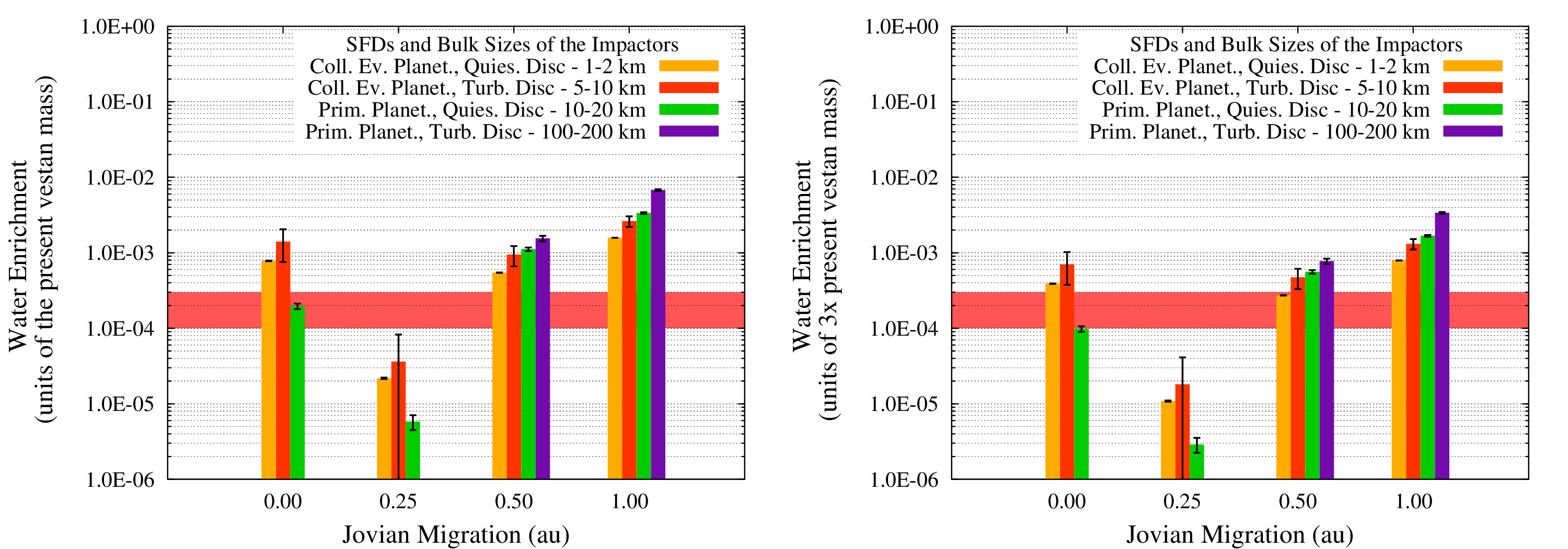}
\caption{Total water accretion experienced by (\emph{left}) a primordial Vesta with the same mass as the present Vesta and (\emph{right}) a primordial Vesta three times as massive during Jupiter's mass growth in the different migration scenarios and for the different SFDs considered. For each SFD we report the characteristic diameter of the planetesimals producing the bulk of the impact flux as computed with our Monte Carlo methods. The horizontal regions highlighted in {\color{black}red} mark the range of values of Vesta's water enrichment {\color{black} and represent our upper boundary to Vesta's water accretion (see Sect. \ref{section-constraints} and \citealt{stephant2016a,stephant2016b,sarafian2017a,sarafian2017b}). Note that, given that the temporal interval considered in this proof-of-concept study is smaller than the timespan over which Vesta's crust can be enriched in water, only those scenarios producing water enrichments \emph{below} the red regions should be considered compatible with present-day Vesta.}}\label{fig-water-total}
\end{figure*}

The cumulative water enrichments produced by asteroidal and cometary impactors in the different migration scenarios for Jupiter are shown in Fig. \ref{fig-water-total}, where they are compared with the range of values for Vesta's water mass fraction derived from the estimates of \citet{stephant2016a,stephant2016b} {\color{black}and \citet{sarafian2017a,sarafian2017b}}. The cases where Jupiter migrated by 0.5 au or more appear inconsistent with the observational data, as the volatile-rich asteroidal impactors would produce a water enrichment from a few times to an order of magnitude larger.

The case of no migration of Jupiter also shows inconsistencies with the observational data, but in this case the inconsistencies appear to be also SFD-dependent. 
Collisionally evolved SFDs produce water enrichments greater than the ranges of values derived from the estimates of \citet{stephant2016a,stephant2016b} {\color{black}and \citet{sarafian2017a,sarafian2017b}} while primordial SFDs are associated to lower ones. In the case of primordial planetesimals formed in quiescent discs the produced water enrichment is just below the range of values derived from eucrites, while in the extreme case of primordial planetesimals formed in a turbulent circumsolar disc no water enrichment is produced (beyond Vesta's initial water budget, if different from zero).

As in the case of mass loss,{\color{black} we tested how these results would change in the ``altered Vesta'' scenario, where primordial Vesta is hypothesized to have been more massive than its present counterpart \citep{consolmagno2015}.} If we consider again a primordial Vesta twice as massive as present Vesta, the increase in the escape velocity should increase the average efficiency of impacts in delivering water by about $5\%$ (see Eq. 8 in \citealt{svetsov2011}). At the same time, the increase in the radius would translate in a proportional increase in the flux of impactors. 

Therefore, a larger primordial Vesta would accrete material more efficiently from a larger number of bodies, partially counteracting the drop in the water enrichment caused by the increase in the crustal mass over which to distribute the accreted water. As a result, the values shown in Figs. \ref{fig-water} and \ref{fig-water-total} would decrease only by about $33\%$ for a primordial Vesta twice as massive as the present one. For a primordial Vesta three times as massive as the present one, the decrease would amount to about $50\%$.

As one can see from Fig. \ref{fig-water-total}, such a decrease does not qualitatively change the outcome of our earlier analysis. Jovian displacements of 0.5 au or larger would still be inconsistent with the constraint posed by the water enrichment of eucrites. Likewise, a lack of migration by Jupiter would be inconsistent with said constraint for collisionally evolved SFDs of the impactors dominated in number by planetesimals smaller than about 10 km (as in the SFDs by \citealt{weidenschilling2011} and \citealt{morbidelli2009}).

\subsection{Mass accretion and HSEs enrichment}

\begin{figure*}[t]
\centering
\includegraphics[width=\textwidth]{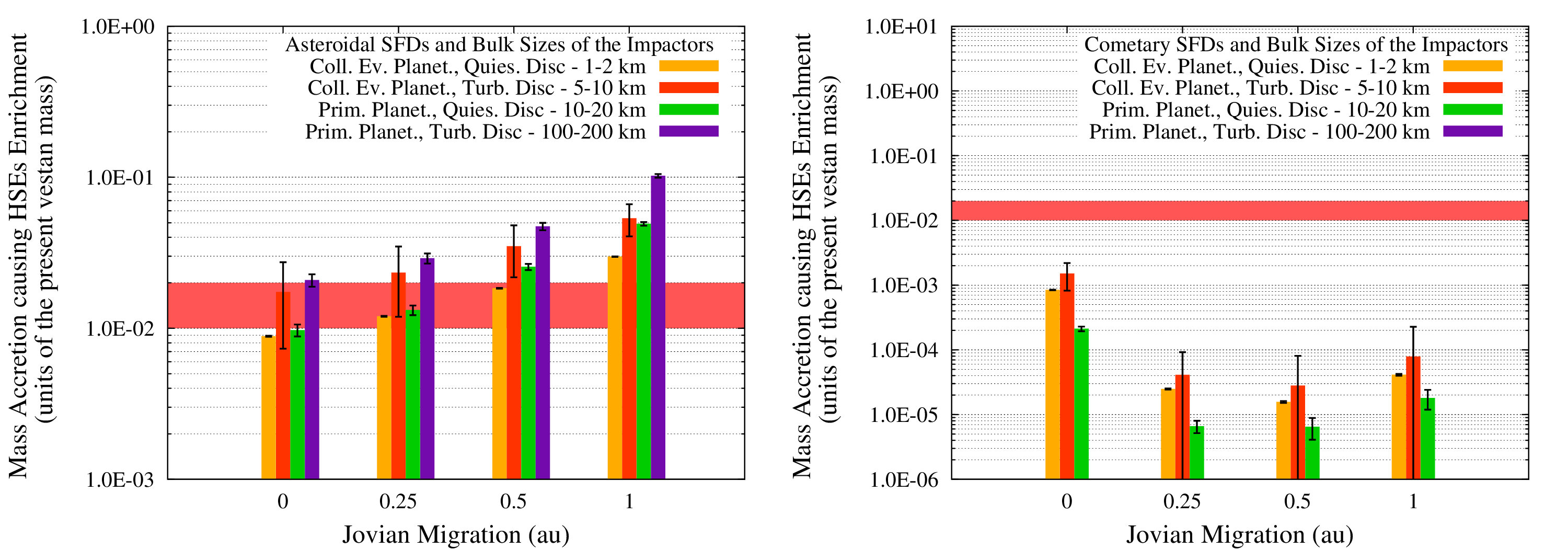}
\caption{Mass accretion responsible for the HSEs enrichment experienced by a primordial Vesta with mass similar to that of the present Vesta due to (\emph{left}) asteroidal impactors and (\emph{right}) cometary impactors during Jupiter's mass growth in the different migration scenarios and for the different SFDs considered. For each SFD we report the characteristic diameter of the planetesimals producing the bulk of the impact flux as computed with our Monte Carlo methods. The horizontal regions highlighted in {\color{black}red} mark the range of values of Vesta's mass accretion needed to produce the observed HSEs enrichment {\color{black} and represent our upper boundary to Vesta's mass accretion (see Sect. \ref{section-constraints} and \citealt{day2012}). Note that, given that the temporal interval considered in this proof-of-concept study is smaller than the timespan over which Vesta's crust can be enriched in HSEs, only those scenarios producing mass accretions \emph{below} the red regions should be considered compatible with present-day Vesta.}}\label{fig-hse}
\end{figure*}

The final step of our analysis was to compare the effects of the global accretion of chondritic material experienced by Vesta with the HSEs enrichment of diogenites, {\color{black} starting also in this case with the classical ``intact and pristine Vesta'' scenario, where the asteroid always possessed a mass similar to its present one}. In computing such accretion we considered, alongside with the contribution of asteroidal impactors, that of the non-ice component of the cometary impactors (see Sect. \ref{section-planetesimals}). The individual contributions of asteroidal and cometary impactors are shown in Fig. \ref{fig-hse}.

The accretion of chondritic material associated to asteroidal impactors (Fig. \ref{fig-hse}, left panel) increases proportionally to Jupiter's displacement due to the growing flux of impactors experienced by Vesta \citep{turrini2011}. The accretion associated to cometary impactors (Fig. \ref{fig-hse}, right panel) follows the same pattern seen when discussing the accretion of water (see Fig. \ref{fig-water}, right panel) and proves marginal with respect to that of asteroidal impactors.

\begin{figure*}[t]
\centering
\includegraphics[width=\textwidth]{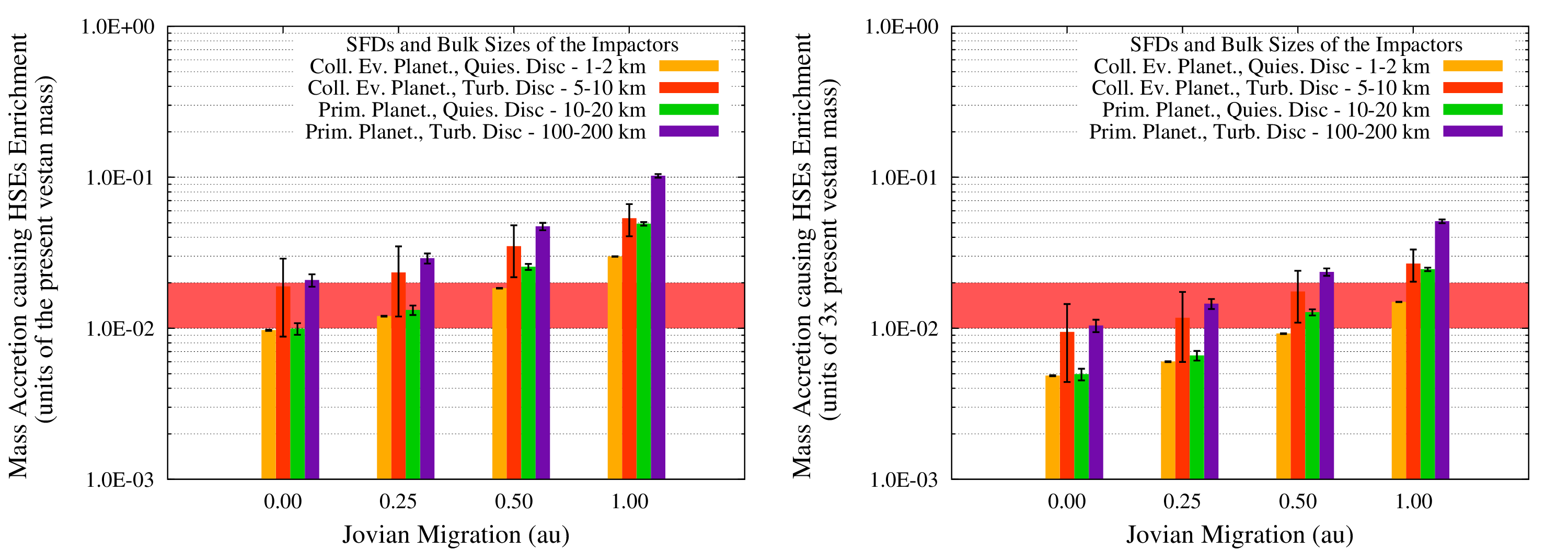}
\caption{Total mass accretion responsible for the HSEs enrichment experienced by (\emph{left}) a primordial Vesta with the same mass as present Vesta and (\emph{right}) a primordial Vesta three times as massive during Jupiter's mass growth in the different migration scenarios and for the different SFDs considered. For each SFD we report the characteristic diameter of the planetesimals producing the bulk of the impact flux as computed with our Monte Carlo methods. The horizontal regions highlighted in {\color{black}red} mark the range of values of Vesta's mass accretion needed to produce the observed HSEs enrichment {\color{black} and represent our upper boundary to Vesta's mass accretion (see Sect. \ref{section-constraints} and \citealt{day2012}). Note that, given that the temporal interval considered in this proof-of-concept study is smaller than the timespan over which Vesta's crust can be enriched in HSEs, only those scenarios producing mass accretions \emph{below} the red regions should be considered compatible with present-day Vesta.}}\label{fig-hse-total}
\end{figure*}

The overall late accretion experienced by Vesta is shown in Fig. \ref{fig-hse-total} and immediately reveals two striking features. The first one is that planetesimals formed in a turbulent circumsolar disc, independently on them being primordial or collisionally evolved, appear to be not consistent with the constraint posed by the HSEs enrichment of diogenites. The second one is that in general the mass accretion experienced by a primordial Vesta with mass similar to that of the present Vesta appears to be at most marginally consistent with said constraint.

In the cases of limited (0.25 au) and no migration, planetesimals formed in quiescent discs produce a mass accretions of about $1\%$ of the vestan mass while those formed in turbulent discs produce a mass accretions of about $2\%$. In the cases of moderate (0.5 au) and large (1 au) migration, the resulting mass accretion is of about $2\%$ of the vestan mass or larger for all kinds of impactors. As we discussed in Sect. \ref{section-constraints}, while \citet{day2012} estimated the accreted mass to fall between $1\%$ and $2\%$ of the mass of Vesta, {\color{black}we treated this range of values as an upper limit in this study to account for the uncertainties on the interpretation of the diogenitic data and for the fact that the process we are considering} lasted only a fraction of the total time over which diogenites can be enriched in HSEs by impacts (see Sect. \ref{section-constraints}).

For a primordial Vesta with a mass similar to the present one of the asteroid, therefore, the cases that best fit the HSEs data among those considered here are those of no or limited (0.25 au) migration of Jupiter in a quiescent circumsolar disc. Even these cases, however, produce an enrichment reaching the lower end of the range identified by \citet{day2012}. We therefore tested the behaviour of the accretion of chondritic mass {\color{black} in the ``altered Vesta'' scenario considering} a primordial Vesta twice or three times larger than the present one.

Applying the same scaling discussed for water accretion to the values shown in Fig. \ref{fig-hse-total}, we can see that a primordial Vesta two to three times more massive than the present Vesta (see Fig. \ref{fig-hse-total}) would make planetesimals formed in turbulent discs (like in the SFDs by \citealt{chambers2010} and \citealt{morbidelli2009}) more consistent with the HSEs constraint in the scenarios of limited (0.25 au) or no migration of Jupiter. At the same time, it would make the case of collisionally evolved planetesimals formed in quiescent discs (like the SFD by \citealt{weidenschilling2011}) more consistent with the HSEs constraint also for a moderate displacement (0.5 au) of Jupiter.

\section{Discussion and conclusions}\label{section-discussion}

{\color{black}The goal we set for ourselves in this work was to investigate whether the erosional and accretional history of the primordial Vesta as recorded by the HEDs can be used to probe into the early collisional history of asteroid Vesta and, through that, into the early evolution of the Solar System. Before discussing the results we obtained, however, we emphasize once again  that they should be considered only as illustrative (or just as a more refined back-of-the-envelope calculation) since some of the approximations adopted in our proof-of-concept case study were motivated only by reasons of convenience and neglected important processes, like gas drag, that should be included in future more physically complete investigations. Because of this, in the following we will limit ourselves to discussing the general trends we observed in our results.}

{\color{black}Notwithstanding its limitations, the proof-of-concept case study we investigated appears to indicate that the three compositional characteristics of Vesta and the HEDs we considered in this work (namely, the survival of Vesta's basaltic crust, the enrichment in water of eucrites and the enrichment in HSEs of diogenites) offer complementary pieces of information that, once considered together, provide stronger constraints than when considered individually. Moreover, the constraints they provide only rely on the assumption of a chondritic bulk composition of Vesta in terms of its major rock-forming elements and, as the comparison between the ``intact and pristine Vesta'' scenario and the ``altered Vesta'' scenario highlights, they appear to be limitedly influenced by the proposed uncertainty on Vesta's primordial mass.

In our proof-of-concept case study the crustal survival to cratering erosion allows to reject only the case of a Jovian migration of 1 au. The constraint offered by the survival of Vesta's basaltic crust to cratering erosion would therefore appear to be the least powerful among those we investigated, as the information it provides is already contained {\color{black}within} that provided by the two constraints associated to late accretion. The accretion history of the primordial Vesta appears instead to provide stronger constraints: both water accretion and mass accretion agree in rejecting the cases of Jovian migration of 0.5 and 1 au, with water accretion also indicating that the case of no migration of the giant planet is inconsistent with the HEDs data, particularly if the D/H ratio of the planetesimal population represented by our cometary impactors was inconsistent with that reported for Vesta's source of water \citep{sarafian2014}.

Among the three constraints, water accretion appears more sensitive to the effects of Jupiter's migration, effectively pinpointing it to about 0.25 au among the simplified cases considered. Mass accretion appears more capable of discriminating between the effects of different size distributions of the impacting planetesimals, favouring the collisionally-evolved SFDs in contrast to primordial ones and the SFDs associated to quiescent nebular environments in contrast to those associated to turbulent nebular environments. Notwithstanding its apparent weakness, the survival of Vesta's basaltic crust remains an important constraint when studying more violent collisional scenarios {\color{black}than} those here considered. 

Specifically, the collisional evolution of the primordial Vesta in those scenarios dominated by high-velocity or even high-energy impacts (e.g. the so-called ``Grand Tack'', \citealt{walsh2011,obrien2014}) will be determined by mass loss without mass accretion playing a significant role. This leading role of mass loss will be particularly true for scenarios invoking a major role of ``hit-and-run'' collisions, like those suggested to be responsible for the ``altered Vesta'' scenario \citep{consolmagno2015}, in the collisional evolution of the inner Solar System, as in those cases the contribution of said impacts to mass accretion will be null or negligible. 

It should be noted, moreover, that in case of stochastic large impacts it is possible for a scenario to be characterized by a moderate or even limited global crustal erosion but a large local excavation. This is indeed the case of the last 4 Gyr of collisional evolution of Vesta, where the total crustal erosion was limited to about 30 m but the impacts that produced Veneneia and Rheasilvia locally excavated tens of km. As proposed in \citet{turrini2011} and further discussed in \citet{turrini2014b} and \citet{turrini2014c}, impacts of this kind occurring on primordial Vesta could cause effusive events where the magma originates from the mantle and could in principle produce compositional signatures in Vesta's crust incompatible with Dawn's measurements. Given the degree of collisional remixing of Vesta's crust suggested by Dawn's observations \citep{desanctis2012,prettyman2012}, these scenarios should be investigated on a case-by-case basis if they can successfully pass the test on the global crustal survival. It is interesting to note, however, that those scenarios that could produce the excavation or effusion of mantle material in \citet{turrini2014b} and \citet{turrini2014c} are among those rejected by the three constraints.

The scenarios we considered in our proof-of-concept case study represent only a limited subset of all proposed evolutionary tracks for the early Solar System. As an example, it has been proposed that Vesta could have formed on an inner orbit located between the orbit of Mars and the inner edge of asteroid belt \citep{bottke2006} instead of in the inner asteroid belt. It is also possible for the giant planets to have undergone a more extensive migration than that considered in this work \citep{walsh2011,bitsch2015}. This extensive migration, in turn, could have kept them in the outer Solar System \citep{bitsch2015} or could have brought them to cross the inner Solar System \citep{walsh2011}. All these different possibilities will be associated to different fluxes of impactors on Vesta and will need to be tested case by case against the three astrochemical constraints we identified.

Also the scenarios we considered for primordial Vesta do not exhaust all the different possibilities. As an example, it has been proposed that a slower formation of Vesta could cause the heat released by the short-lived radioactive elements not to be enough to melt the conductive lid of the asteroid, which would preserve its original undifferentiated composition \citep{formisano2013}. This undifferentiated crust would be reprocessed over time by the effusive processes responsible for the creation of Vesta's basaltic crust, as discussed in Sect. \ref{section-constraints}, and could therefore represent a source of HSEs and possibly water for the vestan magma, whose effects on the enrichment of eucrites and diogenites need to be verified against the astrochemical constraints on Vesta's late accretion.

{\color{black}Finally, the temporal interval covered by our proof-of-concept case study spans only a fraction of the temporal windows (see Sect. \ref{section-constraints}) over which Vesta's crust can be compositionally altered or eroded by impacts: later events, therefore, are also expected to leave their marks on Vesta and the HEDs. In particular, in the scenarios we investigated it is expected that, after Jupiter's formation, the interplay between the gravitational perturbations of the giant planet and those of the planetary embryos embedded into the primordial asteroid belt will start a phase of dynamical excitation and clearing of the belt itself \citep{wetherill1992,petit2001,obrien2007}, changing its orbital structure to its present one (albeit with a larger population of asteroids). Planetesimals impacting Vesta during this phase of dynamical excitation and clearing will also contribute to the mass accretion and mass loss histories of the asteroid and their effects will cumulate with those of the Jovian Early Bombardment.}

Applying the three astrochemical constraints we investigated to a more deterministic study of the history of the early Solar System is beyond the scope of our proof-of-concept case study and is left to future works based on a more complete physical model and spanning longer temporal intervals. In particular, future works will need to include the effects of gas drag, which will change both the flux of impactors on Vesta and the distribution of the impact velocities, and of the population of planetary embryos embedded into the planetesimal disk, which is expected to both dynamically excite the planetesimals and start a process of depletion of the asteroid belt once Jupiter has completed its formation (the latter process becoming more efficient in case of an eccentric orbit of the forming Jupiter), in assessing the collisional evolution of primordial Vesta.

In conclusion, the main result of this work is the identification of the constraints offered by eucrites and diogenites and the showcasing of their joint use as a window into the ancient past of the Solar System. Our take home message can be summarized by the following ``Lather, Rinse, Repeat'' recipe for future studies. Pick the scenario for Vesta that you consider most realistic, put it into the scenario for the evolution of the early Solar System that you want to investigate, and include all the necessary physical ingredients. Let it evolve and check if Vesta's resulting accretional and erosional histories are consistent with the global constraints offered by eucrites and diogenites. Start over as many time as needed.
}

\section*{Acknowledgements}

The authors wish to thank Miroslav Bro{\v z}, an anonymous referee, Chris Russell and the whole Dawn team. This research has been supported by the Italian Space Agency (ASI) and by the International Space Science Institute (ISSI) in Bern through the International Teams 2012 project ``Vesta, the key to the origins of the Solar System'' (\url{www.issibern.ch/teams/originsolsys}). The computational resources used in this research have been supplied by INAF-IAPS through the projects ``HPP - High Performance Planetology'' and  ``DataWell.''

\bibliographystyle{elsarticle-harv}

\begin{thebibliography}{}

\bibitem[Ammannito et al.(2013)]{ammannito2013} Ammannito, E., and 21 colleagues 2013.\ Olivine in an unexpected location on Vesta's surface.\ Nature 504, 122-125. 
\bibitem[Barrett et al.(2016)]{barrett2016} Barrett, T.~J., Barnes, J.~J., Tart{\`e}se, R., Anand, M., Franchi, I.~A., Greenwood, R.~C., Charlier, B.~L.~A., Grady, M.~M.\ 2016.\ The abundance and isotopic composition of water in eucrites.\ Meteoritics and Planetary Science 51, 1110-1124.
\bibitem[Benz and Asphaug(1999)]{benz1999} Benz, W., Asphaug, E.\ 1999.\ Catastrophic Disruptions Revisited.\ Icarus 142, 5-20.
\bibitem[Bitsch et al.(2015)]{bitsch2015} Bitsch B., Lambrechts M., Johansen A.\ 2015. The growth of planets by pebble accretion in evolving protoplanetary discs. Astronomy and Astrophysics 582, article id. A112.
\bibitem[Bizzarro et al.(2005)]{bizzarro2005} Bizzarro, M., Baker, J.~A., Haack, H., Lundgaard, K.~L.\ 2005.\ Rapid Timescales for Accretion and Melting of Differentiated Planetesimals Inferred from $^{26}$Al-$^{26}$Mg Chronometry.\ The Astrophysical Journal 632, L41-L44. 
\bibitem[Bottke et al.(2006)]{bottke2006} Bottke W.~F., Nesvorn{\'y} D., Grimm R.~E., Morbidelli A., O'Brien D.~P.\ 2006.\ Iron meteorites as remnants of planetesimals formed in the terrestrial planet region. Nature 439, 821-824.
\bibitem[Bouvier and Wadhwa(2010)]{bouvier2010} Bouvier, A., Wadhwa, M.\ 2010.\ The age of the Solar System redefined by the oldest Pb-Pb age of a meteoritic inclusion.\ Nature Geoscience 3, 637-641.
\bibitem[Britt et al.(2002)]{britt2002} Britt, D.~T., Yeomans, D., Housen, K., Consolmagno, G.\ 2002.\ Asteroid Density, Porosity, and Structure.\ Asteroids III 485-500. 
\bibitem[Bro{\v z} et al.(2013)]{broz2013} Bro{\v z}, M., Morbidelli, A., Bottke, W.~F., Rozehnal, J., Vokrouhlick{\'y}, D., Nesvorn{\'y}, D.\ 2013.\ Constraining the cometary flux through the asteroid belt during the late heavy bombardment.\ Astronomy and Astrophysics 551, A117. 
\bibitem[Carry(2012)]{carry2012} Carry, B.\ 2012.\ Density of asteroids.\ Planetary and Space Science 73, 98-118.
\bibitem[Chambers(2010)]{chambers2010} Chambers, J.~E.\ 2010.\ Planetesimal formation by turbulent concentration.\ Icarus 208, 505-517. 
\bibitem[Clenet et al.(2014)]{clenet2014} Clenet, H., Jutzi, M., Barrat, J.-A., Asphaug, E.~I., Benz, W., Gillet, P.\ 2014.\ A deep crust-mantle boundary in the asteroid 4 Vesta.\ Nature 511, 303-306.
\bibitem[Consolmagno and Drake(1977)]{consolmagno1977} Consolmagno, G.~J., Drake, M.~J.\ 1977.\ Composition and evolution of the eucrite parent body - Evidence from rare earth elements.\ Geochimica et Cosmochimica Acta 41, 1271-1282. 
\bibitem[Consolmagno et al.(2015)]{consolmagno2015} Consolmagno, G.~J., Golabek, G.~J., Turrini, D., Jutzi, M., Sirono, S., Svetsov, V., Tsiganis, K.\ 2015.\ Is Vesta an intact and pristine protoplanet?.\ Icarus 254, 190-201.
\bibitem[Consolmagno et al.(2016)]{consolmagno2016} Consolmagno G.~J., Rubie D.~C. and Golabek G.~J.,\ 2016. The grand tack, Vesta, and the missing olivine problem. Meteoritical Society annual meeting abstract n. 6066.
\bibitem[Coradini et al.(1981)]{coradini1981} Coradini, A., Magni, G., Federico, C.\ 1981.\ Formation of planetesimals in an evolving protoplanetary disk.\ Astronomy and Astrophysics 98, 173-185. 
\bibitem[Coradini, Magni, \& Turrini(2010)]{coradini2010} Coradini, A., Magni, G., Turrini, D.\ From gas to satellitesimals: Disk formation and evolution.\ Space Sci. Rev. 153, 411-429.
\bibitem[Coradini et al.(2011)]{coradini2011} Coradini, A., Turrini, D., Federico, C., Magni, G.\ 2011.\ Vesta and Ceres: Crossing the History of the Solar System.\ Space Science Reviews 163, 25-40.
\bibitem[Cuzzi et al.(2008)]{cuzzi2008} Cuzzi, J.~N., Hogan, R.~C., Shariff, K.\ 2008.\ Toward Planetesimals: Dense Chondrule Clumps in the Protoplanetary Nebula.\ The Astrophysical Journal 687, 1432-1447. 
\bibitem[Cuzzi et al.(2010)]{cuzzi2010} Cuzzi, J.~N., Hogan, R.~C., Bottke, W.~F.\ 2010.\ Towards initial mass functions for asteroids and Kuiper Belt Objects.\ Icarus 208, 518-538. 
\bibitem[D'Angelo, Durisen \& Lissauer(2011)]{dangelo2011} D'Angelo G., Durisen R.~H., Lissauer J.~J. Giant Planet Formation. In {\em Exoplanets}; edited by S. Seager. University of Arizona Press: Tucson, Arizona, 2011, pp. 319-346.
\bibitem[Dale et al.(2012)]{dale2012} Dale, C.~W., Burton, K.~W., Greenwood, R.~C., Gannoun, A., Wade, J., Wood, B.~J., Pearson, D.~G.\ 2012.\ Late Accretion on the Earliest Planetesimals Revealed by the Highly Siderophile Elements.\ Science 336, 72. 
\bibitem[Davis et al.(1979)]{davis1979} Davis, D.~R., Chapman, C.~R., Greenberg, R., Weidenschilling, S.~J., Harris, A.~W.\ 1979.\ Collisional evolution of asteroids - Populations, rotations, and velocities.\ Asteroids 528-557. 
\bibitem[Davis et al.(1985)]{davis1985} Davis, D.~R., Chapman, C.~R., Weidenschilling, S.~J., Greenberg, R.\ 1985.\ Collisional history of asteroids: Evidence from Vesta and the Hirayama families.\ Icarus 62, 30-53. 
\bibitem[Day et al.(2012)]{day2012} Day, J.~M.~D., Walker, R.~J., Qin, L., Rumble, D., III 2012.\ Late accretion as a natural consequence of planetary growth.\ Nature Geoscience 5, 614-617. 
\bibitem[Day et al.(2016)]{day2016} Day J.~M.~D., Brandon A. D., Walker, R.~J., 2016. Highly Siderophile Elements in Earth, Mars, the Moon, and Asteroids. Reviews in Mineralogy $\&$ Geochemistry 81, 161-238
\bibitem[Daly and Schultz(2016)]{daly2016} Daly, R.T., Schultz, P.H., 2016. Delivering a projectile component to the vestan regolith. Icarus 264, 9-19.
\bibitem[De Sanctis et al.(2012)]{desanctis2012} De Sanctis, M.~C., and 22 colleagues\ 2012.\ Spectroscopic Characterization of Mineralogy and Its Diversity Across Vesta.\ Science 336, 697.
\bibitem[DeMeo and Carry(2014)]{demeo2014} DeMeo, F.~E., Carry, B.\ 2014.\ Solar System evolution from compositional mapping of the asteroid belt.\ Nature 505, 629-634. 
\bibitem[Dhaliwal et al.(2016)]{dhaliwal2016} Dhaliwal, J.~K., Day, J.~M.~D., Tait, K.~T.\ 2016.\ Establishing a Pristinity Index for Eucrites Using the Highly Siderophile Elements.\ Lunar and Planetary Science Conference 47, 2644.
\bibitem[Dienes and Walsh(1970)]{dienes1970} Dienes, J.K.; Walsh, J.M. Theory of Impact: Some General Principles and the Method of Eulerian Codes. In {\em High-Velocity Impact Phenonena};  Kinslow, R., Ed.; Academic Press: New York, NY, USA, 1970; pp. 46--104.
\bibitem[Ermakov et al.(2014)]{ermakov2014} Ermakov, A.~I., Zuber, M.~T., Smith, D.~E., Raymond, C.~A., Balmino, G., Fu, R.~R., Ivanov, B.~A.\ 2014.\ Constraints on Vesta's interior structure using gravity and shape models from the Dawn mission.\ Icarus 240, 146-160. 
\bibitem[Lodders(2010)]{lodders2010} Lodders K. \ 2010.\ Solar System Abundances of the Elements. In: Goswami A., Reddy B. (eds) Principles and Perspectives in Cosmochemistry. Astrophysics and Space Science Proceedings 16. Springer, Berlin, Heidelberg, pp. 379-417.
\bibitem[Fedele et al.(2010)]{fedele2010} Fedele D., van den Ancker M. E., Henning Th., Jayawardhana R., Oliveira J. M., 2010. Timescale of mass accretion in pre-main-sequence stars. Astronomy and Astrophysics 510, id. A72. 
\bibitem[Formisano et al.(2013)]{formisano2013} Formisano, M., Federico, C., Turrini, D., Coradini, A., Capaccioni, F., De Sanctis, M.~C., Pauselli, C.\ 2013.\ The heating history of Vesta and the onset of differentiation.\ Meteoritics and Planetary Science 48, 2316-2332.
\bibitem[Goldreich and Ward(1973)]{goldreich1973} Goldreich, P., Ward, W.~R.\ 1973.\ The Formation of Planetesimals.\ The Astrophysical Journal 183, 1051-1062. 
\bibitem[Grazier et al.(2014)]{grazier2014} Grazier, K.~R., Castillo-Rogez, J.~C., Sharp, P.~W.\ 2014.\ Dynamical delivery of volatiles to the outer main belt.\ Icarus 232, 13-21.
\bibitem[Greenwood et al.(2014)]{greenwood2014} Greenwood, R.~C., Barrat, J.-A., Yamaguchi, A., Franchi, I.~A., Scott, E.~R.~D., Bottke, W.~F., Gibson, J.~M.\ 2014.\ The oxygen isotope composition of diogenites: Evidence for early global melting on a single, compositionally diverse, HED parent body.\ Earth and Planetary Science Letters 390, 165-174. 
\bibitem[Hartogh et al.(2011)]{hartogh2011} Hartogh, P., Lis, D.C., Bockel{\'e}e-Morvan, D., de Val-Borro, M., Biver, N., K{\"u}ppers, M., Emprechtinger, M., Bergin, E.A., Crovisier, J., Rengel, M., et al (2011). Ocean-like water in the Jupiter-family comet 103P/Hartley 2. Nature 478, 218--220.
\bibitem[Holsapple(1993)]{holsapple1993} Holsapple, K.~A.\ 1993.\ The scaling of impact processes in planetary sciences.\ Annual Review of Earth and Planetary Sciences 21, 333-373.
\bibitem[Holsapple and Housen(2007)]{holsapple2007} Holsapple, K.~A., Housen, K.~R.\ 2007.\ A crater and its ejecta: An interpretation of Deep Impact.\ Icarus 187, 345-356. 
\bibitem[Ivanov and Melosh(2013)]{ivanov2013} Ivanov, B.~A., Melosh, H.~J., 2013. Two-dimensional numerical modeling of the Rheasilvia impact formation. J. Geophys. Res. 118, 1545-1557. 
\bibitem[Jarosewich(1990)]{jarosewich1990} Jarosewich, E.\ 1990.\ Chemical analyses of meteorites - A compilation of stony and iron meteorite analyses.\ Meteoritics 25, 323-337. 
\bibitem[Johnson et al.(2016)]{johnson2016} Johnson B.~C., Walsh K.~J., Minton D.~A., Krot A.~N., Levison H.~L. (2016). Timing of the formation and migration of giant planets
as constrained by CB chondrites. Science Advances 2, art. id. e1601658, DOI: 10.1126/sciadv.1601658.
\bibitem[Jutzi et al.(2013)]{jutzi2013} Jutzi, M., Asphaug, E., Gillet, P., Barrat, J.-A., Benz, W.\ 2013.\ The structure of the asteroid 4 Vesta as revealed by models of planet-scale collisions.\ Nature 494, 207-210.
\bibitem[Kruijer et al.(2017)]{kruijer2017} Kruijer, T.~S., Burkhardt, C., Budde, G., Kleine, T.\ 2017.\ Age of Jupiter inferred from the distinct genetics and formation times of meteorites.\ Proceedings of the National Academy of Science 114, 6712-6716. 
\bibitem[Lissauer et al.(2009)]{lissauer2009} Lissauer, J.J., Hubickyj, O., D'Angelo, G., Bodenheimer, P.\ Models of Jupiter's growth incorporating thermal and hydrodynamic constraints.\ Icarus 199, 338-350.
\bibitem[Mandler and Elkins-Tanton(2013)]{mandler2013} Mandler, B.~E., Elkins-Tanton, L.~T.\ 2013.\ The origin of eucrites, diogenites, and olivine diogenites: Magma ocean crystallization and shallow magma chamber processes on Vesta.\ Meteoritics and Planetary Science 48, 2333-2349. 
\bibitem[McCord et al.(1970)]{mccord1970} McCord, T.~B., Adams, J.~B., Johnson, T.~V.\ 1970.\ Asteroid Vesta: Spectral Reflectivity and Compositional Implications.\ Science 168, 1445-1447.
\bibitem[McSween et al.(2011)]{mcsween2011} McSween H.~Y., Mittlefehldt D.~W., Beck A.~W., Mayne R.~G., and McCoy T. J. 2011. HED meteorites and their relationship to the geology of Vesta and the Dawn Mission. Space Science Reviews 163, 141-174
\bibitem[Melosh(1989)]{melosh1989} Melosh, H.J., 1989.\ Impact Cratering: A Geologic Process; Oxford Monographs on Geology and Geophysics, No.~11; Oxford University Press: New York,  NY, USA; p. 253.
\bibitem[Michalak(2000)]{michalak2000} Michalak, G.\ 2000.\ Determination of asteroid masses --- I. (1) Ceres, (2) Pallas and (4) Vesta.\ Astronomy and Astrophysics 360, 363-374. 
\bibitem[Michtchenko et al.(2016)]{michtchenko2016} Michtchenko, T.~A., Lazzaro, D., Carvano, J.~M.\ 2016.\ On the current distribution of main belt objects: Constraints for evolutionary models.\ Astronomy and Astrophysics 588, A11. 
\bibitem[Moore et al.(2017)]{moore2017} Moore W.~B., Simon J.~I., Alexander A., Webb G. (2017). Heat-pipes planets. Earth and Planetary Science Letters 474, 13-19.
\bibitem[Morbidelli et al.(2009)]{morbidelli2009} Morbidelli, A., Bottke, W.~F., Nesvorn{\'y}, D., Levison, H.~F.\ 2009.\ Asteroids were born big.\ Icarus 204, 558-573. 
\bibitem[Morbidelli and Raymond(2016)]{morbidelli2016} Morbidelli A., Raymond S.N.,\ 2016. Challenges in planet formation. J. Geophys. Res. Planets 121, 1962-1980.
\bibitem[O'Brien, Morbidelli \& Bottke(2007)]{obrien2007} O'Brien D.~P., Morbidelli A., Bottke W.~F., 2007. The primordial excitation and clearing of the asteroid belt - Revisited.\ Icarus 191, 434-452
\bibitem[O'Brien and Sykes(2011)]{obrien2011} O'Brien, D.~P., Sykes, M.~V.\ 2011.\ The Origin and Evolution of the Asteroid Belt - Implications for Vesta and Ceres.\ Space Science Reviews 163, 41-61.
\bibitem[O'Brien et al.(2014)]{obrien2014} O'Brien, D.~P., Walsh, K.~J., Morbidelli, A., Raymond, S.~N., Mandell, A.~M.\ 2014.\ Water delivery and giant impacts in the Grand Tack scenario.\ Icarus 239, 74-84. 
\bibitem[{\"O}pik(1976)]{opik1976} {\"O}pik, E.~J.\ 1976.\ Interplanetary encounters: close-range gravitational interactions. Developments in solar system and space science (Amsterdam (Netherlands): Elsevier Scientific Publishing), 2, 7 + 155 p.\ .
\bibitem[Petit, Morbidelli \& Chambers(2001)]{petit2001} Petit J.~M., Morbidelli A., Chambers J.~E., 2001.\ The Primordial Excitation and Clearing of the Asteroid Belt.\ Icarus 153, 338-347. 
\bibitem[Pierazzo et al.(1997)]{pierazzo1997} Pierazzo, E.; Vickery, A.M.; Melosh, H.J., 1997. A reevaluation of impact melt production. Icarus 127, 408--423.
\bibitem[Pirani and Turrini(2016)]{pirani2016} Pirani, S., Turrini, D.\ 2016.\ Asteroid 4 Vesta: Dynamical and collisional evolution during the Late Heavy Bombardment.\ Icarus 271, 170-179.
\bibitem[Prettyman et al.(2012)]{prettyman2012} Prettyman, T.~H., and 19 colleagues 2012.\ Elemental Mapping by Dawn Reveals Exogenic H in Vesta's Regolith.\ Science 338, 242.
\bibitem[Raymond \& Izidoro(2017)]{raymond2017} Raymond, S.~N., \& Izidoro, A., 2017. Origin of water in the inner Solar System: Planetesimals scattered inward during Jupiter and Saturn's rapid gas accretion. Icarus 297, 134-148
\bibitem[Robert(2003)]{robert2003} Robert, F.\ 2003.\ The D/H 
Ratio in Chondrites.\ Space Science Reviews 106, 87-101.
\bibitem[Roszjar et al.(2016)]{roszjar2016} Roszjar J., et al., 2016. Prolonged magmatism on 4 Vesta inferred from Hf-W analyses of eucrite zircon. Earth and Planetary Science Letters 452, 216-226.
\bibitem[Ruesch et al.(2014)]{ruesch2014} Ruesch, O., and 14 colleagues 2014.\ Detections and geologic context of local enrichments in olivine on Vesta with VIR/Dawn data.\ Journal of Geophysical Research (Planets) 119, 2078-2108. 
\bibitem[Russell et al.(2012)]{russell2012} Russell, C.~T., and 27 colleagues 2012.\ Dawn at Vesta: Testing the Protoplanetary Paradigm.\ Science 336, 684.
\bibitem[Russell et al.(2013)]{russell2013} Russell, C.~T., and 23 colleagues 2013.\ Dawn completes its mission at 4 Vesta.\ Meteoritics and Planetary Science 48, 2076-2089.
\bibitem[Safronov(1969)]{safronov1969} Safronov, V.~S.\ Evolution of the protoplanetary cloud and formation of the earth and planets. Nauka Press, 1969. Translated from Russian.~Jerusalem (Israel): Israel Program for Scientific Translations, Keter Publishing House, 1972, pp. 212.
\bibitem[Sarafian et al.(2013)]{sarafian2013} Sarafian, A.~R., Roden, M.~F., Pati{\~n}o-Douce, A.~E.\ 2013.\ The volatile content of Vesta: Clues from apatite in eucrites.\ Meteoritics and Planetary Science 48, 2135-2154.
\bibitem[Sarafian et al.(2014)]{sarafian2014} Sarafian, A.~R., Nielsen, S.~G., Marschall, H.~R., McCubbin, F.~M., Monteleone, B.~D.\ 2014.\ Early accretion of water in the inner solar system from a carbonaceous chondrite-like source.\ Science 346, 623-626. 
\bibitem[Sarafian et al.(2017a)]{sarafian2017a} Sarafian, A.~R., Nielsen, S.~G., Marschall, H.~R., Gaetani, G.~A., Hauri, E.~H., Righter, K., Berger, E.~L.\ 2017.\ Volatile Concentrations and H-Isotope Composition of Unequilibrated Eucrites.\ Lunar and Planetary Science Conference 48, 1436.
\bibitem[Sarafian et al.(2017b)]{sarafian2017b} Sarafian, A.~R., John, T., Roszjar, J., Whitehouse, M.~J.\ 2017.\ Chlorine and hydrogen degassing in Vesta's magma ocean.\ Earth and Planetary Science Letters 459, 311-319. 
\bibitem[Schenk et al.(2012)]{schenk2012} Schenk, P., and 13 colleagues\ 2012.The Geologically Recent Giant Impact Basins at Vesta's South Pole.\ Science 336, 694. 
\bibitem[Scott(2006)]{scott2006} Scott, E.~R.~D.\ Meteoritical and dynamical constraints on the growth mechanisms and formation times of asteroids and Jupiter.\ Icarus 185, 72-82. 
\bibitem[Scott(2007)]{scott2007} Scott, E.~R.~D.\ 2007.\ Chondrites and the Protoplanetary Disk.\ Annual Review of Earth and Planetary Sciences 35, 577-620. 
\bibitem[Schiller et al.(2011)]{schiller2011} Schiller, M., Baker, J., Creech, J., Paton, C., Millet, M.-A., Irving, A., Bizzarro, M.\ 2011.\ Rapid Timescales for Magma Ocean Crystallization on the Howardite-Eucrite-Diogenite Parent Body.\ The Astrophysical Journal 740, L22. 
\bibitem[Shuvalov(1999)]{shuvalov1999} Shuvalov, V.V. (1999) Multi-dimensional hydrodynamic code SOVA for interfacial flows: Application to thermal layer effect. Shock Waves 9, 381-390.
\bibitem[Stephant et al.(2016a)]{stephant2016a} Stephant, A., Hervig, R.~L., Wadhwa, M.\ 2016.\ Water in Nominally Anhydrous Crustal Minerals of Vesta.\ Lunar and Planetary Science Conference 47, 2436.
\bibitem[Stephant et al.(2016b)]{stephant2016b} Stephant, A., Hervig, R., Bose, M., Wadhwa, M.\ 2016.\ D/H Ratios and Water Contents in Eucrite Minerals: Implications for the Source and Abundance of Water on Vesta.\ LPI Contributions 1921, 6212.
\bibitem[Steenstra et al.(2016)]{steenstra2016} Steenstra, E.~S., Knibbe, J.~S., Rai, N., van Westrenen, W.\ 2016.\ Constraints on core formation in Vesta from metal-silicate partitioning of siderophile elements.\ Geochimica et Cosmochimica Acta 177, 48-61. 
\bibitem[Svetsov(2011)]{svetsov2011} Svetsov, V.\ 2011.\ Cratering erosion of planetary embryos.\ Icarus 214, 316-326. 
\bibitem[Svetsov and Shuvalov(2015)]{svetsov2015} Svetsov, V.~V., Shuvalov, V.~V.\ 2015.\ Water delivery to the Moon by asteroidal and cometary impacts.\ Planetary and Space Science 117, 444-452. 
\bibitem[Thomas et al.(1997)]{thomas1997} Thomas, P.C., Binzel, R.P., Gaffey, M.J., et al.,\ 1997. Vesta: Spin pole, size, and shape from HST images. Icarus 128, 88-94.
\bibitem[Thompson and Lauson(1972)]{thompson1972} Thompson, S.L.; Lauson, H.S., 1972. Improvements in the Chart D Radiation-Hydrodynamic CODE III: Revised Analytic Equations of State; Report SC-RR-71 0714; Sandia National Laboratory: Albuquerque, NM, USA; p. 119.
\bibitem[Tillotson(1962)]{tillotson1962} Tillotson, J.H., 1962. Metallic Equations of State for Hypervelocity Impact; General Atomic Report GA-3216; Advanced Research Project Agency: San Diego, CA, USA; p. 140.
\bibitem[Toplis et al.(2013)]{toplis2013} Toplis, M.~J., and 10 colleagues 2013.\ Chondritic models of 4 Vesta: Implications for geochemical and geophysical properties.\ Meteoritics and Planetary Science 48, 2300-2315. 
\bibitem[Tkalcec et al.(2013)]{tkalcec2013} Tkalcec, B.~J., Golabek	G. J., Brenker F.~E.\ 2013.\ Solid-state plastic deformation in the dynamic interior of a differentiated asteroid.\ Nature Geoscience 6, 93-97.
\bibitem[Turrini et al.(2011)]{turrini2011} Turrini, D., Magni, G., Coradini, A.\ 2011.\ Probing the history of Solar system through the cratering records on Vesta and Ceres.\ Monthly Notices of the Royal Astronomical Society 413, 2439-2466.
\bibitem[Turrini et al.(2012)]{turrini2012} Turrini, D., Coradini, A., Magni, G.\ 2012.\ Jovian Early Bombardment: Planetesimal Erosion in the Inner Asteroid Belt.\ The Astrophysical Journal 750, id. 8. 
\bibitem[Turrini(2014)]{turrini2014b} Turrini, D.\ 2014.\ The primordial collisional history of Vesta: crater saturation, surface evolution and survival of the basaltic crust.\ Planetary and Space Science 103, 82-95. 
\bibitem[Turrini \& Svetsov(2014)]{turrini2014c} Turrini, D., Svetsov, V.\ 2014.\ The Formation of Jupiter, the Jovian Early Bombardment and the Delivery of Water to the Asteroid Belt: The Case of (4) Vesta.\ Life 4, 4-34. 
\bibitem[Turrini et al.(2014)]{turrini2014d} Turrini, D., and 12 colleagues 2014.\ The contamination of the surface of Vesta by impacts and the delivery of the dark material.\ Icarus 240, 86-102. 
\bibitem[Turrini et al.(2015)]{turrini2015} Turrini, D., Nelson, R.~P., Barbieri, M.\ 2015.\ The role of planetary formation and evolution in shaping the composition of exoplanetary atmospheres.\ Experimental Astronomy 40, 501-522. 
\bibitem[Turrini et al.(2016)]{turrini2016} Turrini, D., Svetsov, V., Consolmagno, G., Sirono, S., Pirani, S.\ 2016.\ Olivine on Vesta as exogenous contaminants brought by impacts: Constraints from modeling Vesta's collisional history and from impact simulations.\ Icarus 280, 328-339.
\bibitem[Walsh et al.(2011)]{walsh2011} Walsh, K.~J., Morbidelli, A., Raymond, S.~N., O'Brien, D.~P., Mandell, A.~M.\ 2011.\ A low mass for Mars from Jupiter's early gas-driven migration.\ Nature 475, 206-209.
\bibitem[Wang et al.(2017)]{wang2017} Wang, H., Weiss, B.~P., Bai, X.-N., Downey, B.~G., Wang, J., Wang, J., Suavet, C., Fu, R.~R., Zucolotto, M.~E.\ 2017.\ Lifetime of the solar nebula constrained by meteorite paleomagnetism.\ Science 355, 623-627. 
\bibitem[Weidenschilling(1975)]{weidenschilling1975} Weidenschilling, S.~J. 1975.\ Mass loss from the region of Mars and the asteroid belt.\ Icarus 26, 361-366. 
\bibitem[Weidenschilling(1980)]{weidenschilling1980} Weidenschilling, S.~J.\ 1980.\ Dust to planetesimals - Settling and coagulation in the solar nebula.\ Icarus 44, 172-189.
\bibitem[Weidenschilling, Davis \& Marzari(2001)]{weidenschilling2001} Weidenschilling, S.~J., Davis, D.~R., Marzari, F. 2001.\ Very early collisional evolution in the asteroid belt.\ Earth, Planets, and Space 53, 1093-1097. 
\bibitem[Weidenschilling(2008)]{weidenschilling2008} Weidenschilling, S.~J.\ 2008.\ Accretion of planetary embryos in the inner and outer solar system.\ Physica Scripta Volume T 130, 014021.
\bibitem[Weidenschilling(2011)]{weidenschilling2011} Weidenschilling, S.~J.\ 2011.\ Initial sizes of planetesimals and accretion of the asteroids.\ Icarus 214, 671-684. 
\bibitem[Wetherill(1992)]{wetherill1992} Wetherill G. W., 1992.\ An alternative model for the formation of asteroids.\ Icarus 100, 307-325




\end{thebibliography}



\end{document}